\documentclass[conference,compsoc]{IEEEtran}
\IEEEoverridecommandlockouts

\usepackage{amsmath,amssymb}

\usepackage{pifont} 

\usepackage{bbding} 
\newcommand{\cmark}{\textcolor{green!80!black}{\ding{51}}}
\newcommand{\xmark}{\textcolor{purple}{\ding{55}}}

\usepackage{xcolor}
\definecolor{ao(english)}{rgb}{0.0, 0.5, 0.0}

\usepackage{makecell}
\usepackage{colortbl} 

\usepackage{tikz}
\usetikzlibrary{arrows.meta}
\usetikzlibrary{automata,positioning}
\usepackage{placeins}

\usepackage{hyperref}
\hypersetup{
    colorlinks=true,
    linkcolor=blue,
    filecolor=magenta,      
    urlcolor=black,
    citecolor = black,
    pdftitle={An Example},
    pdfpagemode=FullScreen,
    }
\usepackage{threeparttable}

\renewcommand*{\arraystretch}{1.5}%
\definecolor{tabred}{RGB}{230,36,0}%
\definecolor{tabgreen}{RGB}{0,116,21}%
\definecolor{taborange}{RGB}{250,124,30}%
\definecolor{tabbrown}{RGB}{171,70,0}%
\definecolor{tabyellow}{RGB}{235,240,120}

\definecolor{palepeach}{RGB}{240,210,170}  
\definecolor{softblue}{RGB}{170,180,240}   
\definecolor{mintwhite}{RGB}{210,230,210} 
\newcommand*{\vcorr}{%
  \vadjust{\vspace{-\dp\csname @arstrutbox\endcsname}}%
  \global\let\vcorr\relax
}%

\usepackage{mathtools}

\usepackage{lscape}
\usepackage[figuresright]{rotating}
\usepackage[graphicx]{realboxes}
\usepackage{adjustbox}
\usepackage{caption}

\usepackage{subfigure}

\usepackage{colortbl} 
\usepackage{xfrac}

\usepackage{float}

\usepackage{stackengine}
\usepackage{scalerel}
\usepackage{xcolor}
\newcommand\dangersign[1][2ex]{%
  \renewcommand\stacktype{L}%
  \scaleto{\stackon[1.3pt]{\color{red}$\triangle$}{\tiny !}}{#1}%
}

\newcommand{\qw}[1]{\textcolor{blue}{#1}}
\newcommand{\mfcomment}[1]{\textcolor{purple}{#1}}

\usepackage{fbox} 
\usepackage{fancybox}
\usepackage{framed}

\usepackage{multirow}

\def\BibTeX{{\rm B\kern-.05em{\sc i\kern-.025em b}\kern-.08em
    T\kern-.1667em\lower.7ex\hbox{E}\kern-.125emX}}
    
\usepackage{CJKutf8}  
\usepackage{pgfplots}
\usepackage{filecontents}

\usepackage{tabularx,makecell, caption}
\usepackage{enumitem} 

\newcolumntype{L}{>{\arraybackslash}X}

\usepackage{multicol}
\usepackage{listings}
\usepackage{blindtext}
\usepackage{etoolbox,xstring,mfirstuc,textcase}
\usepackage{listings}

\usepackage{subfiles}
\usepackage[most]{tcolorbox}

\usepackage{xcolor}
\usepackage{siunitx}
\usepackage{comment}
\usepackage{indentfirst} 
\usepackage{framed} 

\usepackage[font=small,labelfont=bf,tableposition=top]{caption}
\usepackage{booktabs}
\usepackage{dashbox}
\newcounter{myfootnote}

\usepackage{listings}
\usepackage{url}

\newcommand{\hlhref}[2]{\href{#1}{\textcolor{teal}{{#2}}}}

\usepackage{hyperref}
\hypersetup{
     colorlinks = true,
     linkcolor = blue,
     anchorcolor = red,
     citecolor = magenta,
     filecolor = blue,
     urlcolor = black,
}
\usepackage{color}
\usepackage{algorithm,algpseudocode} 

\usepackage{tikz}
\usepackage{pgfplots}

\definecolor{findOptimalPartition}{HTML}{D7191C}
\definecolor{storeClusterComponent}{HTML}{FDAE61}
\definecolor{dbscan}{HTML}{ABDDA4}
\definecolor{constructCluster}{HTML}{2B83BA}

\newenvironment{packeditemize}{
	\begin{list}{$\bullet$}{
			\setlength{\labelwidth}{4pt}
			\setlength{\itemsep}{0pt}
			\setlength{\leftmargin}{\labelwidth}
			\addtolength{\leftmargin}{\labelsep}
			\setlength{\parindent}{0pt}
			\setlength{\listparindent}{\parindent}
			\setlength{\parsep}{0pt}
			\setlength{\topsep}{1pt}}}{\end{list}}

\begin{document}
\title{BRC20 Pinning Attack}


\author{Minfeng Qi$^{1,\textcolor{green}{*}}$\thanks{\quad $^{\textcolor{green}{*}}$\text{Equal contribution.}}, Qin Wang$^{2,\textcolor{green}{*}}$, Zhipeng Wang$^3$, Lin Zhong$^4$, Zhixiong Gao$^5$,\\ Tianqing Zhu$^1$,  Shiping Chen$^2$,  William Knottenbelt$^3$
~\\
$^1$\textit{City University of Macau} $|$
$^3$\textit{Imperial College London} \\ 
$^2$\textit{CSIRO Data61}  $|$
$^4$\textit{Bitlayer Labs} $|$
$^5$\textit{Binance}
}

\maketitle






\begin{abstract}

BRC20 tokens are a type of non-fungible asset on the Bitcoin network. They allow users to embed customised content within Bitcoin’s satoshis. The token frenzy reached a market size of US\$2,811b (2023Q3-2025Q1). However, this intuitive design has not undergone serious security scrutiny. 

We present the first analysis of BRC20's $\textit{transfer}$ mechanism and identify a new attack vector. A typical BRC20 transfer involves two ``bundled'' on-chain transactions with different fee levels: the first (i.e., \textit{\textbf{Tx1}}) with a lower fee inscribes the $\mathsf{transfer}$ request, while the second (i.e., \textit{\textbf{Tx2}}) with a higher fee finalizes the actual transfer. An adversary can send a manipulated fee transaction (falling between the two fee levels), which makes \textit{\textbf{Tx1}} processed while \textit{\textbf{Tx2}} pinned in the mempool. This locks the BRC20 liquidity and disrupts normal withdraw requests from users. We term this \textit{BRC20 pinning attack}.

We validated the attack in real-world settings in collaboration with Binance researchers. With their knowledge and permission, we conducted a controlled test against Binance’s ORDI hot wallet, resulting in a temporary suspension of ORDI withdrawals for 3.5 hours. Recovery was performed soon. Further analysis confirms that the attack can be applied to over \textbf{90\%} of inscription-based tokens within the Bitcoin ecosystem.

\end{abstract}


\section{Introduction}

BRC20, short for Bitcoin Request for Comment 20 \cite{binance2}\cite{binance3}, is a framework that enables the creation of \textit{non-fungible} assets (or NFTs)~\cite{wang2021non} on the Bitcoin network. 
The core aspect of non-fungibility is to implement uniqueness. Ethereum NFTs~\cite{wang2021non}\cite{wood2014ethereum} introduce $\mathsf{tokenID}$ (defined in ERC721~\cite{erc721}) as the unique identifier for each token. These tokens are exclusively associated (i.e., linked to specific addresses) with external digital assets. Ownership is account-based and transferred via smart contracts. 
In contrast, BRC20~\cite{brc20experiment} is fundamentally different.

BRC20 tokens inscribe additional context directly into each satoshi (\textit{Sat}, the smallest unit\footnote{In the Bitcoin system~\cite{nakamoto2008bitcoin}, 1 BTC equals $10^8$ \textit{Sat}s.})~\cite{wang2023understanding} and assign them sequential identifiers (so that unique) using Ordinal Theory~\cite{ordinaltheory2024}\cite{bipord}. The transfer of ownership is determined by direct transaction movement. Although using plaintext strings, BRC20 inscriptions can be converted into more formats, including text, images, videos, or even opcodes that serve as operational guidance~\cite{binance6} (technical details in Sec.\ref{sec-brc20}).

To date, BRC20 has garnered significant attention, evident from various Bitcoin's market indicators\footnote{$\mathsf{YChart}$ \url{https://ycharts.com/indicators/bitcoin_average_block_size} and \url{https://ycharts.com/indicators/bitcoin_average_transaction_fee} [Oct. 2024];
$\mathsf{BRC20.IO}$ \url{https://www.BRC20.io/}  [Oct. 2024];
$\mathsf{Blockchain.com}$ \url{https://www.blockchain.com/explorer/charts/mempool-count} [Aug. 2023].} including block size (surging from an average of 1.11MB to 1.85MB+), mempool transactions (reaching a peak of 315,257 transactions), and transaction fees (averaging a 10\%+ increase\footnote{Based on data spanning from early Feb. 2023 to Apr. 2024.}). As of early 2025, the total number of BRC20 tokens has reached 795, with a trading volume of US\$83b and a market capitalization of US\$2,811b. Statistical resources (e.g., Dune Analytics~\cite{duneanalytic}\cite{duneanalytic1}) corroborate the upward trend in minted Ordinals.


In parallel, the development of BRC20 was also documented in recent studies (Sec.\ref{sec-rw2}), including a series of reports \cite{binance2}\cite{binance6}\cite{binance1}\cite{binance4}\cite{binance5}, market investigations~\cite{wang2023understanding}\cite{li2024bitcoin}, analytical works~\cite{bertucci2024bitcoin}\cite{szydlo2024characteristics}\cite{wen2024modular}\cite{wang2024bridging}. Many projects with similar design principles are deployed on their corresponding platforms, such as Ethscription~\cite{ethscription24}, BSC20~\cite{bnbchaininscriptions2024}, SOLs~\cite{magicedenunlocking2024}.

Our work differs from those prior ones that simply describe the BRC20 protocols or solely explore auxiliary components (e.g., indexers). Instead, we focus on its core operational transferring mechanisms and present the first security assessment (more solutions compared in Table~\ref{tab:attack_taxonomy}.)

We find that BRC20 is surprisingly vulnerable. It is susceptible to a straightforward attack by freezing one of the transactions (typically needs two) during its $\mathsf{transfer}$ operation. We introduce this attack as \textit{BRC20 pinning attack}. Our attack can cause prolonged functional blockages and can be applied to all relevant hot wallets. Unlike mere programming flaws, the issue stems from an inherent design dilemma within BRC20, making it extremely challenging to fully eliminate. We present our contributions step by step. 

\begin{figure}[!]
\subfigure[\textbf{BRC20's $\mathsf{transfer}$ operation on-chain}: It involves two transactions. \textit{\textbf{Tx1}} inscribes the JSON-style content (i.e., ordi's $\mathsf{transfer}$ operation) into a satoshi, marked with a unique sequence number (e.g., Inscription \#3), and sends it to oneself. \textit{\textbf{Tx2}} executes the actual transfer, carrying Inscription \#3. 
]{\label{fig:transfer}
\begin{minipage}[b]{\linewidth}
\centering
\includegraphics[width=1\linewidth]{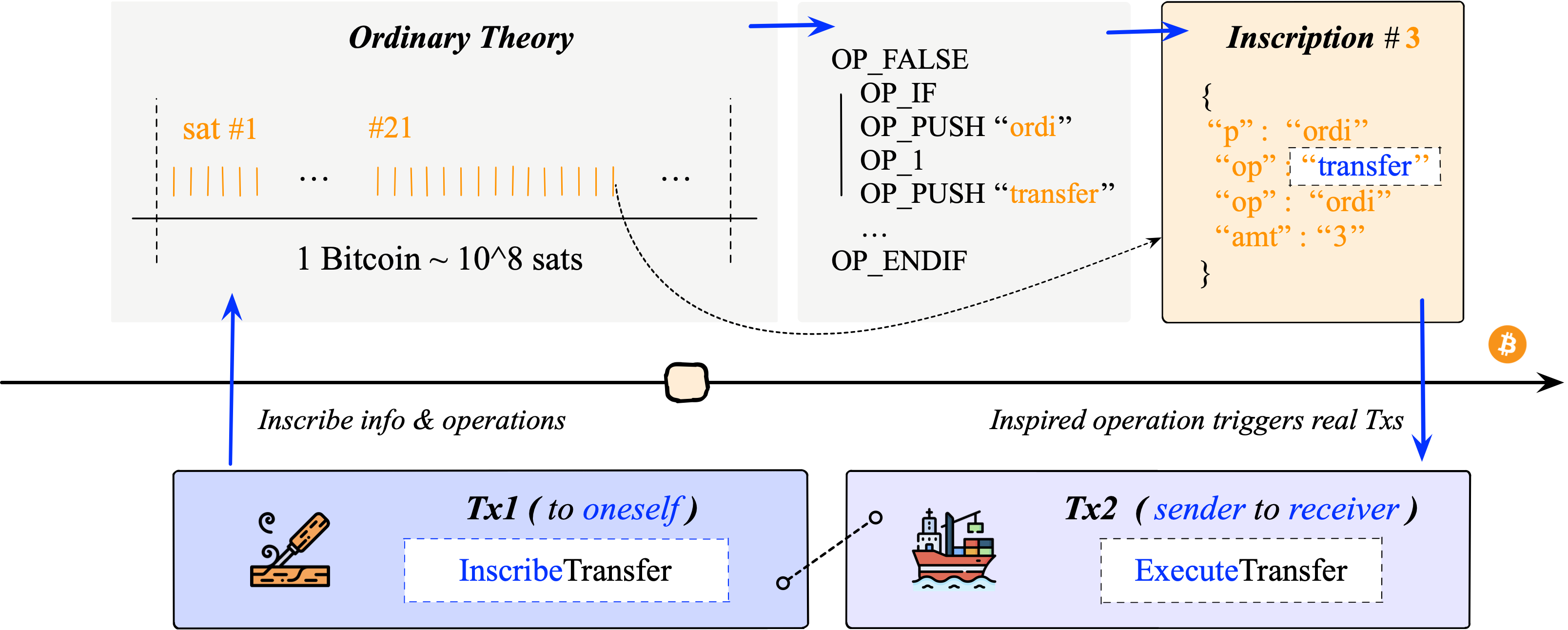}
\end{minipage}
}
\subfigure[\textbf{Transferring inscriptions from end-to-end}: Transferring an inscription is conducted by standard UTXO transactions. A sender sends 8 sats (Inscriptions \#1-\#8) to two recipients: 3 sats (\#1-\#3) are sent to A, 2 sats (\#4, \#5) to B, 2 sats (\#6, \#7) are returned to the sender, and 1 sat (\#8) is used for fees.]{\label{fig:onchain}
\begin{minipage}[b]{\linewidth}
\centering
\includegraphics[width=1\linewidth]{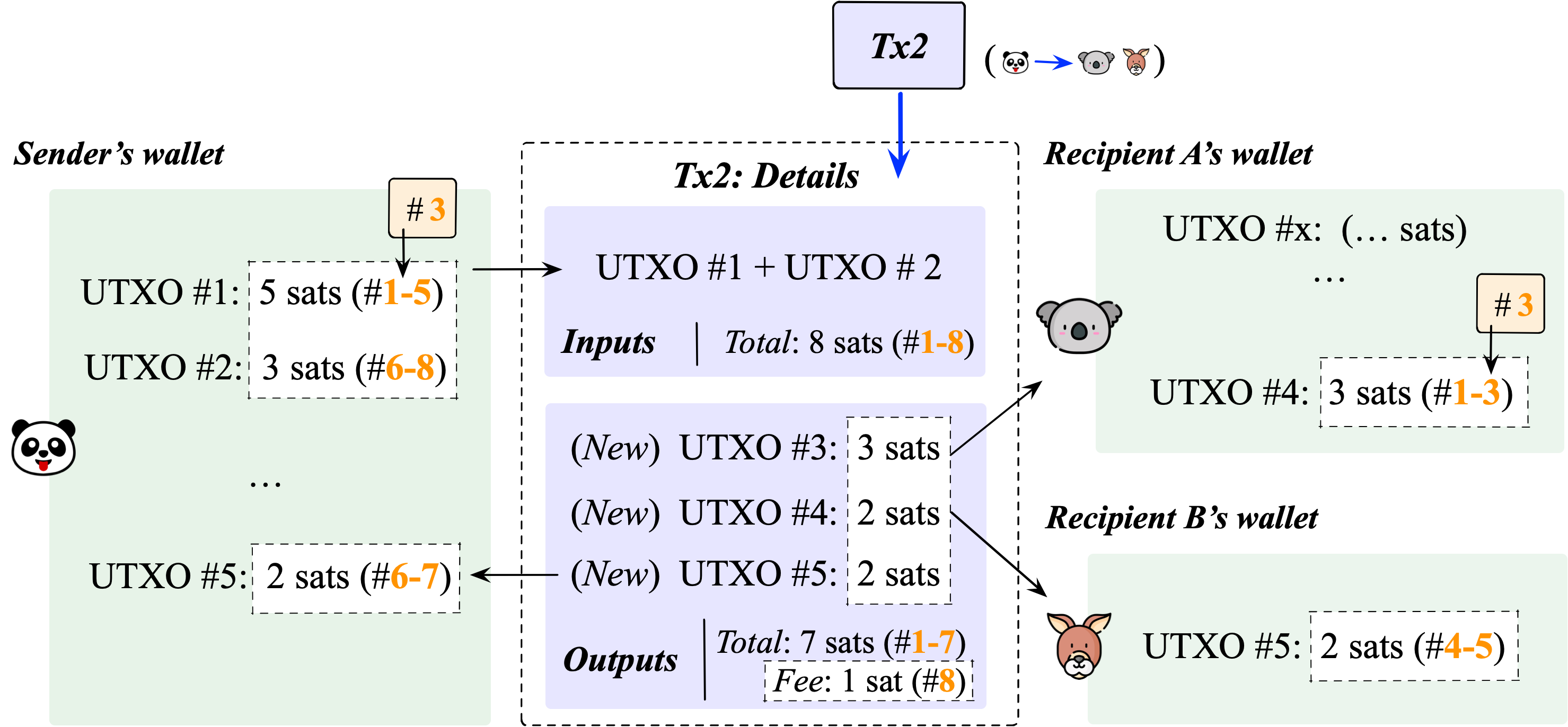}
\end{minipage}
}
\subfigure[\textbf{Off-chain updates for inscribed BRC20 tokens}: The sender's holding BRC20 tokens (ordi) move from an available state to a transferable state after \textit{\textbf{Tx1}}. These tokens are then sent to the recipient after \textit{\textbf{Tx2}}. The recipient's balance increases correspondingly, but only after \textit{\textbf{Tx2}} is completed.
]{ \label{fig:offchain}
\begin{minipage}[b]{\linewidth}
\centering
\includegraphics[width=1\linewidth]{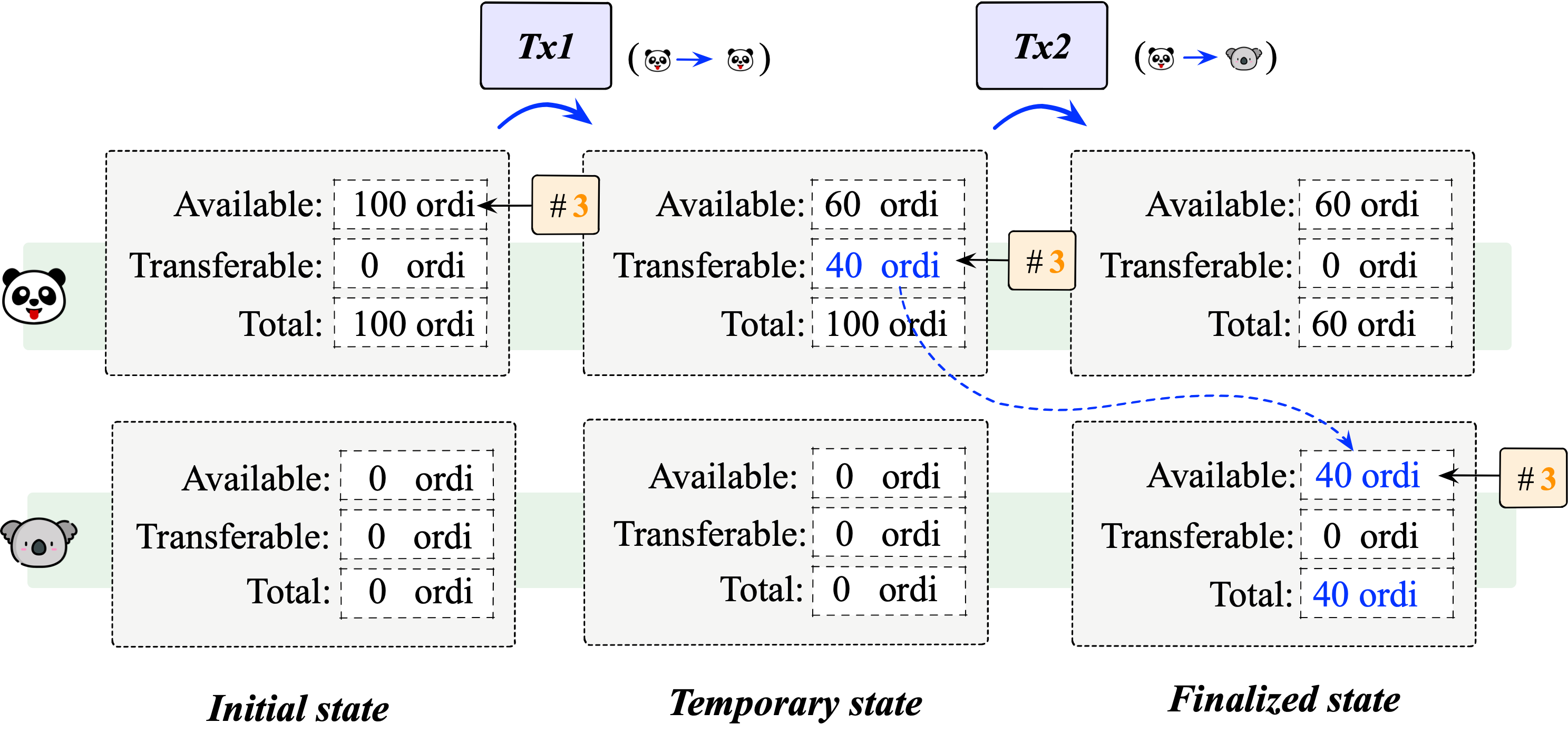}
\end{minipage}
}

\caption{How to transfer BRC20 tokens}
\label{fig-howtransfer}
\vspace{-0.2in}
\end{figure}

\smallskip
{\textbf{\ding{172} Dynamically transferring BRC20s}} (Sec.\ref{sec-transferBRC}, Fig.\ref{fig-howtransfer}).
\smallskip

We provide the first in-depth analysis on \textit{how to transfer BRC20 tokens} over the Bitcoin network. While prior work~\cite{binance2,binance3,wang2023understanding} has primarily examined the \textit{static} aspects of BRC2, such as token inscription and deployment, little attention has been given to the actual mechanics of token transfer across the network. Some recent efforts~\cite{wen2024modular,wang2024bridging} acknowledge the possibility of BRC20 token transfers, but they abstract away the on-chain execution details and instead emphasize the role of off-chain indexers. In contrast, our work systematically analyzes the two-phase transaction model that underpins BRC20 transfers and uncovers critical vulnerabilities in this process.

We find that transferring a BRC20 token is totally different to transferring BTCs. Contrary to what might be recognized, the inscribed $\mathsf{transfer}$ is not an actual operation executed on the Bitcoin network. Instead, it is merely a form of script, written after $\mathsf{OP\_False}$ (implying that it will never be executed on-chain) and stored on-chain. 

The execution of such a $\mathsf{transfer}$ operation relies on the movement of real Bitcoin transactions (technically referred to as UTXOs\footnote{A UTXO (Unspent Transaction Output)~\cite{nakamoto2008bitcoin}\cite{antonopoulos2014mastering} is an unspent output from a Bitcoin transaction that can be used as input in a future transaction.}). To complete a BRC20 $\mathsf{transfer}$ operation, \underline{two} UTXO-based transactions (i.e., \textit{\textbf{Tx1}}, \textit{\textbf{Tx2}} in sequence) are required. \textit{\textbf{Tx1}} inscribes the operation (typically by sending it to oneself) and includes essential information such as the amount and recipient. \textit{\textbf{Tx2}} transfers the inscription to the recipient. These two transactions are typically initiated from an off-chain wallet and ``bundled'' together for execution.

\begin{figure}[t]
\includegraphics[width=1\linewidth]{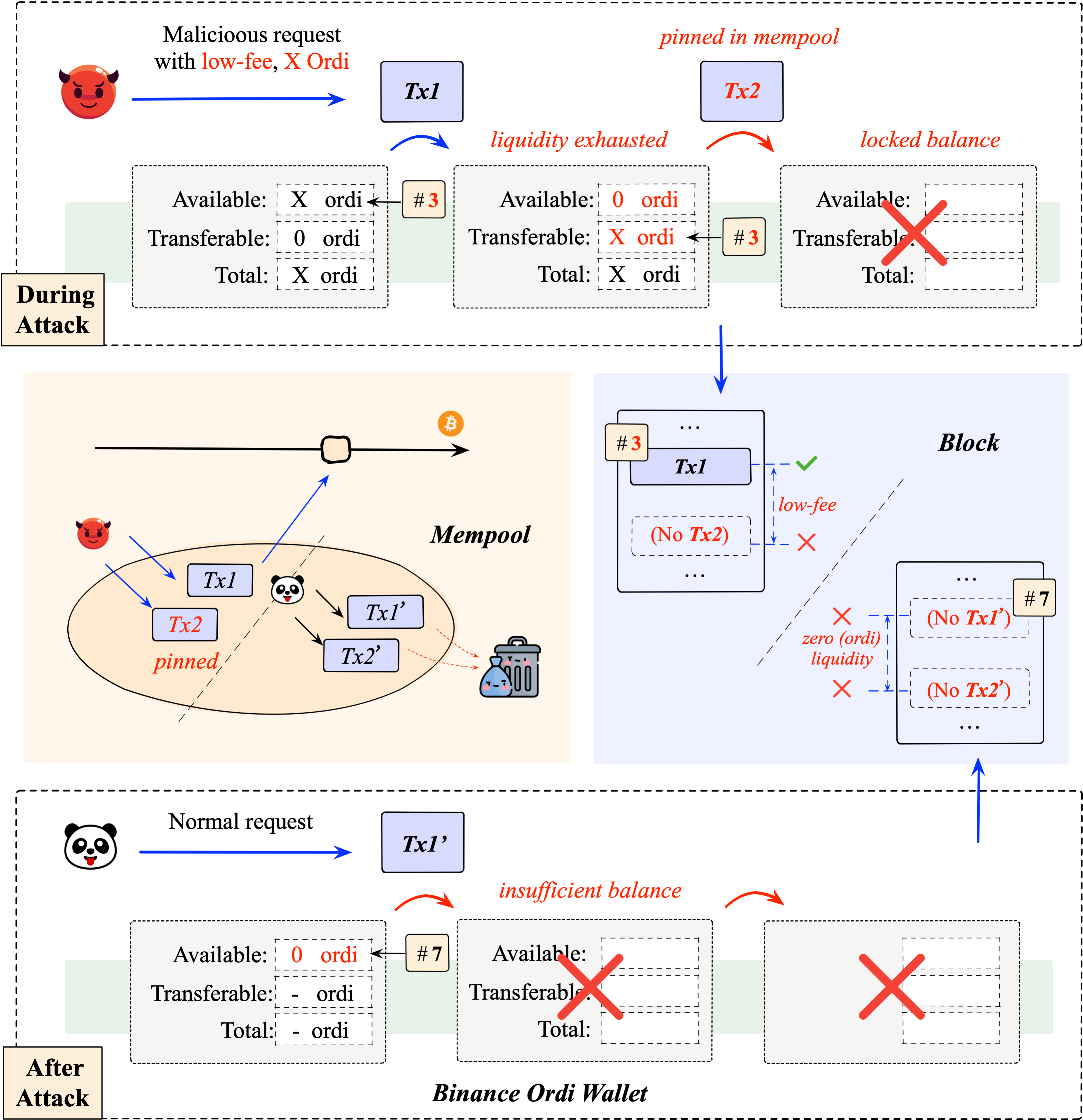}
\caption{\textbf{Our BRC20 pinning attack:} An attacker sends a low-fee transaction (i.e., $f_{\text{tx}}$, where $f_{\text{tx}}\in[f_{\text{min}}, f_{\text{sf}}]$, requesting the withdrawal of the entire ordi balance (i.e., X) from the targeted wallet. \textbf{\textit{Tx1}} is completed since $f_{\text{tx}} > f_{\text{min}}$, moving ordi into a transferable (pending) state. However, \textbf{\textit{Tx2}} cannot be executed as $ f_{\text{tx}} < f_{\text{sf}} $. \textbf{\textit{Tx2}} is pinned in the mempool and locks all ordi liquidity in the transferable state. If a user attempts the subsequent request, even \textbf{\textit{Tx1'}} will fail, as the available balance in the hot wallet is insufficient. Both \textbf{\textit{Tx1'}} and \textbf{\textit{Tx2'}} will be dropped.}
\label{fig-attackflow}
\end{figure}

\smallskip
\textbf{{\ding{173} New attack: BRC20 pinning attack}} (Sec.\ref{sec-pinningAttack}, Fig.\ref{fig-attackflow}). 
\vspace{1pt}

We introduce \textit{BRC20 pinning attack}, a new attack (to our knowledge, also the first) targeted BRC20 tokens. Our attack is rooted in leveraging the BRC20 transfer operation (Sec.\ref{subsec-transferNetwork}) and transaction pinning techniques (Sec.\ref{subsec-txpinning}). 

We discovered that the “bundled” transactions sent from users’ wallets are not actually bundled. The two transactions are created and sent independently, with an imperceptibly short delay that makes them appear as a single bundled transaction. Our attack exploits this gap by pinning \textit{\textbf{Tx2}} in the mempool, delaying the token transfer to the recipient.

To achieve this, we strategically set a low-fee transaction (i.e., the malicious \textit{\textbf{Tx2}}). The fee is carefully chosen to fall between two thresholds: sufficient for \textit{\textbf{Tx1}} to be processed but insufficient for \textit{\textbf{Tx2}}. Since accurate transaction fees are theoritically unknown\footnote{\begin{minipage}[t]{0.45\textwidth}
but what is known: based on empirical investigation, we observed that the fee required for \textit{\textbf{Tx1}} is greatly lower — around 3 to 5 times less — than the fee required for \textit{\textbf{Tx2}}.
\end{minipage}}, we begin with a very low fee (insufficient for either \textit{\textbf{Tx1}} or \textit{\textbf{Tx2}}) and incrementally increase it through repeated rebroadcasts. Once the fee is adequate for \textit{\textbf{Tx1}} but still below the threshold for \textit{\textbf{Tx2}}, our attack succeeds. At this point, \textit{\textbf{Tx2}} will not be picked up by miners immediately and will remain in the mempool for an extended period (becoming \textit{pinned}), locking the associated transferable balance. 
These processes occur on-chain.


For off-chain processes, BRC20 movement relies on the usable state of \textit{transferable balance} (Fig.\ref{fig:offchain}). When a transfer request is initiated, tokens in \textit{\textbf{Tx1}} move from the \textit{available balance} (which shows the current balance but cannot be spent directly) to the \textit{transferable balance} (ready for spending). In \textit{\textbf{Tx2}}, these transferable tokens are sent to the recipient, leaving the \textit{transferable balance} at zero. In this sense, our attack immobilizes liquidity by prolonging the time tokens are locked within the transferable balance.



\smallskip
{\textbf{\ding{174} Launching controlled attack validation} }(Sec.\ref{sec-localExperi}\&\ref{sec-attackBinance}).

We first implemented and evaluated the pinning behavior in a local, controlled environment. A series of experiments were conducted to analyze how various factors (including transfer amount, transaction fee, mempool congestion level, and the number of transaction attempts) affect the likelihood and persistence of the pinning condition.

Our evaluation reveals that the frequency of attack attempts, combined with low transaction fees, can substantially extend transaction confirmation delays, especially under high mempool congestion. Moreover, our results indicate that both larger individual transfers and a greater number of small transfers can effectively lock liquidity in a target’s transferable balance.

Building on these local experiments, we conducted a controlled real-world validation in collaboration with Binance’s research team. With their full awareness and permission, we tested the attack against Binance’s ORDI hot wallet, the most popular and active BRC20 wallet (handling 26.24\%~\cite{ordimarket24} of the total transaction volume). Through four carefully monitored attack attempts, we successfully locked the entire transferable balance of the wallet. 

\vspace{0.05in}
\begin{center}
\colorbox{teal!8}{
\begin{minipage}{0.93\linewidth}
\small
\textbf{Responsible actions and results.} We jointly monitor transaction status with Binance. The attack resulted in a suspension of ORDI withdrawals for around 3.5 hours. Recovery transactions were soon initiated to restore normal wallet operations.
\end{minipage}
} 
\end{center}

\smallskip
\textbf{\ding{175} Extended discussion} (Sec.\ref{sec-discussion}).
\smallskip

To address the risks exposed by our findings, we further investigate defense strategies against BRC20 pinning attacks. In particular, we propose and evaluate several mitigation techniques, including dynamic fee adjustment and real-time transaction monitoring. Our controlled experiments show that adjusting minimum fee thresholds in accordance with mempool congestion levels can raise the cost of sustaining long-lived unconfirmed transactions, thereby limiting the attack surface. In addition, real-time detection of anomalous behavior, such as unusually large transfer amounts submitted with abnormally low fees, can serve as a useful early warning mechanism for exchanges and wallet providers.

However, despite these countermeasures, the underlying design of the BRC20 protocol presents inherent limitations. Since BRC20 lacks programmability (e.g., conditional logic or access control), distinguishing malicious operations from legitimate ones at the protocol level is challenging. Defense can only be detection-based, rather than preventive.

\vspace{-0.05in}
\begin{center}
\colorbox{magenta!11}{
\begin{minipage}{0.95\linewidth}
\small
\textbf{Attack (and downstream) impact.} The testing confirmed that the attack does not structurally compromise the BRC20 protocol, but can temporarily disable key wallet functions (e.g., withdrawals). Our attack will not cause direct financial loss. However, it introduces measurable opportunity costs, particularly for upper-layer DeFi activities reliant on timely liquidity.
\end{minipage}
} 
\vspace{0.05in}
\end{center}

We also discuss the adaptability of our attack to other BRC20 tokens and various inscription protocols on other blockchains. Our findings show that the attack is broadly applicable to \textbf{90\%}+ of BRC20 tokens on Bitcoin, which follow the same standardized inscription-based transfer mechanism as ORDI. However, the attack proves \underline{in}effective (in around \textbf{80\%} of cases) against inscription protocols, such as ETHS20 on Ethereum and TON20 on Arbitrum. This is primarily because these protocols utilize a single-transaction process rather than a two-step transfer method, eliminating the possibility of delaying a second transaction.

\section{Dissecting BRC20}
\label{sec-brc20}



\subsection{Technical Foundation}

\noindent\textbf{Bitcoin ``standards''.} 
Bitcoin standards are primarily defined through Bitcoin Improvement Proposals (BIPs)~\cite{bips}, which serve as formal specifications governing various components of the Bitcoin protocol and its broader ecosystem. Similar in purpose to Ethereum's ERC standards, BIPs cover a wide range of topics, including transaction formats, address schemes, wallet structures, and security mechanisms. In Table~\ref{tab:target_node_state}, we highlight the subset of BIPs that are particularly relevant to the design and creation of BRC20 tokens.

\smallskip
\noindent\textbf{Schnorr signature.} The Schnorr signature scheme~\cite{schnorr1991efficient}, formalized in BIP-340, was introduced to replace ECDSA in Bitcoin, offering improved efficiency and comparable security guarantees. In particular, Schnorr signatures reduce the signature size from 71–72 bytes (ECDSA) to a fixed 64 bytes. One of its key advantages is support for key aggregation (also known as \textit{multisig}), which enables multiple parties to collaboratively generate a single aggregated signature.

\smallskip
\noindent\textbf{Taproot/Tapscript.} Taproot (BIP-341/342/343) outlines the integration of Schnorr signatures and introduces the concept of locking outputs to multiple scripts using Merkle branches (MAST, BIP-114/117). This enables a new method of spending Bitcoin known as Pay-to-Taproot (P2TR) where users can pay to either a Schnorr public key or the Merkle root of various other scripts  (e.g., Tapscript).

\smallskip
\noindent\textbf{Segregated witness.} SegWit (BIP-141/144) enables larger scripting space. It introduces the \textit{witness} structure that can host extra data. A SegWit transaction has two parts: the base transaction (containing inputs and outputs, excluding signatures) and the witness section (appended afterward, carrying  signature and script data). SegWit reduces transaction storage size and increases block capacity from 1MB to 4MB.

\smallskip
\noindent\textbf{Mempool transaction exit.}
In Bitcoin, unconfirmed transactions remain in the mempool but are removed under specific conditions. (i) The most common scenario is when a transaction is confirmed by being included in a block mined by a miner. Once confirmed, the transaction is removed from the mempool, as it no longer needs to be broadcasted. (ii) Additionally, if the mempool reaches its maximum capacity, lower-fee transactions may be evicted to make space for higher-fee ones. (iii) Moreover, transactions can also exit the mempool through the RBF mechanism~\cite{bitcoinRBF}, where a user broadcasts a new transaction with a higher fee, replacing the original lower-fee transaction. (iv) Another reason for transaction removal is expiration. If a transaction remains unconfirmed in the mempool for an extended period — usually 14 days~\cite{imtiaz2019churn} — it may be dropped by nodes due to timeout. (v) Finally, conflicting transactions, such as those involved in double-spend attempts, will be removed if a competing transaction is prioritized.

\begin{table}[!]
    \caption{BRC20-related BIPs} \label{tab:target_node_state}
    \label{Tab}
      \centering
      \renewcommand{\arraystretch}{1} 
        \resizebox{\linewidth}{!}{
        \begin{tabular}{c|c|c}
            \toprule
            \multicolumn{1}{c}{\textbf{BIP-}} &  \multicolumn{1}{c}{\textbf{Title}} &  \multicolumn{1}{c}{Usage for BRC20} \\
            \midrule
             \hlhref{https://github.com/bitcoin/bips/blob/master/bip-0114.mediawiki}{114}  & \cellcolor{gray!10} Merkelized Abstract Syntax Tree (MAST) & Batch Validation/Processing  \\

              \hlhref{https://github.com/bitcoin/bips/blob/master/bip-0117.mediawiki}{117}  & \cellcolor{gray!10} Tail Call Execution Semantics  & Batch Validation/Processing \\

              \hlhref{https://github.com/bitcoin/bips/blob/master/bip-0125.mediawiki}{125}  &  \cellcolor{gray!10} Opt-in Full Replace-by-Fee Signaling  & Transaction Pinning \\
              
             \hlhref{https://github.com/bitcoin/bips/blob/master/bip-0141.mediawiki}{141} & \cellcolor{gray!10} Segregated Witness (Consensus layer) & Space Extension \\
             
             \hlhref{https://github.com/bitcoin/bips/blob/master/bip-0144.mediawiki}{144}  & \cellcolor{gray!10} Segregated Witness (Peer Services) & Space Extension  \\

             \hlhref{https://github.com/bitcoin/bips/blob/master/bip-0340.mediawiki}{340}  & \cellcolor{gray!10} Schnorr Signatures for Secp256k1 & Batch Validation   \\
                         
             \hlhref{https://github.com/bitcoin/bips/blob/master/bip-0341.mediawiki}{341} &  \cellcolor{gray!10} Taproot: SegWit version 1 spending rules & Inscription \\
            
            \hlhref{https://github.com/bitcoin/bips/blob/master/bip-0342.mediawiki}{342} & \cellcolor{gray!10} Validation of Taproot Scripts & Inscription \\
 
            \hlhref{https://github.com/bitcoin/bips/blob/master/bip-0343.mediawiki}{343} & \cellcolor{gray!10} Mandatory activation of Taproot deployment & Inscription  \\
        \bottomrule
        \end{tabular}
       }
\end{table}

\subsection{BRC20 Formation}
\label{brc20-foundation}

\noindent\textbf{Creating uniqueness.} Unlike ERC721, which assigns a $\mathsf{tokenID}$ (an additional customizable field capable of accommodating any number) for NFTs~\cite{wang2021non}, uniqueness in the Bitcoin network is established based on its smallest unit, \textit{satoshi}, via the Ordinal Theory~\cite{ordinaltheory2024}. As satoshis are indivisible, the protocol essentially serves as a numbering scheme for satoshis, enabling each satoshi to be tracked/transferred as an individual entity (cf.~Fig.\ref{fig:transfer}). It is preserved as transactions transfer from inputs to outputs, following the first-in-first-out principle, which is a fundamental feature of UTXO-based transactions. 
We defer the details in Sec.\ref{subsec-transferNetwork}.

\smallskip
\noindent\textbf{Inscribing context.} 
Inscription involves embedding personalized content into each unique satoshi. Such data is assigned the P2TR script type, and the generated inscriptions are stored within the SegWit section of a transaction (\textit{witness}), presented in various formats such as text (JSON formats), images, audio, and even highly compressed video files (peep at \cite{ordinalsbot}\cite{ordiscan}).

We take a text-based BRC20 (e.g., ORDI~\cite{ordicoin}) as the instance. The on-chain inscription contains several crucial details (see Listing~\ref{list-transfer}), including the protocol name ($\mathsf{brc20}$), operation (e.g., $\mathsf{transfer}$), token name (e.g., $\mathsf{ordi}$), total amount of tokens to be issued ($\mathsf{max}$), maximum amount of tokens to be minted in each round ($\mathsf{lim}$), and amount of tokens to be minted/transferred ($\mathsf{amt}$).

\begin{lstlisting}[caption={Inscriptions and the \textcolor{black}{$\mathsf{Transfer}$ operation}}, label={list-transfer}, basicstyle=\ttfamily\scriptsize]
# On-chain Inscription
"p" : "brc20" # protocol name
"op": "deploy" # operation
"tick": "ordi" # token name
"max": "2100000" # total amount of issued token
"lim": "1000" # maximum minted tokens per round
"amt": "100" # the amount of token being actioned

# Inscription with the Transfer operation
"p" : "ordi", # protocol
"op": "transfer", # inscribed transfer operation
"tick": "ordi", # token to be transferred
"amt": "100" # the amount of transferred tokens

# UTXO Transactions over the Bitcoin network

# Off-chain update
if state["tick"] NOT exists OR 
                 "amt" > "lim" OR sum("amt") > "max":
    raise errors
else 
    account["tick"]["balance"][minter] += amt
\end{lstlisting}

\smallskip
\noindent\textbf{On-chain storage.} BRC20 tokens are settled on-chain, and their movement (i.e., implementing the $\mathsf{Transfer}$ operation) depends on the transfer of Bitcoin native transactions (i.e., the sending and receiving of satoshis) via the underlying layer-one network (cf. Fig.\ref{fig:offchain}).
The Bitcoin network only records native transaction activities related to these inscriptions. It does not provide real-time insights into the status of BRC20 balances (unlike account-based models~\cite{wood2014ethereum}). These functionalities are delegated to off-chain layers.

\smallskip
\noindent\textbf{Off-chain retrieves.} Users interact with those on-chain tokens via off-chain validation services (also known as \textit{indexers}~\cite{wang2023understanding}), which facilitate real-time data retrieval. Indexers are responsible for tracking which wallets minted the original token supply up to the maximum limit, determining where minting ceased, and tracing whether tokens traded in secondary markets can be linked back to wallets.

Indexers only track transactions without modifying or appending on-chain data. Even if an indexer fails, there is no risk of fund loss, as it can be deterministically reconstructed by reapplying the transaction rules. Many users may be unaware of these indexers, as their services are often integrated into user-friendly wallets such as Ordinals Wallet~\cite{ordinalwallet}, UniSat~\cite{unisat}, Binance's hot wallet~\cite{binanceweb3wallet}, etc.



\section{Transferring BRC20 Tokens}
\label{sec-transferBRC}

\subsection{Inscribed $\mathsf{Transfer}$ Operation }

\noindent\textbf{BRC20 Operations.} The BRC20 protocol defines three core operations: (i) $\mathsf{deploy}$: used to create a new BRC20 token, allowing the issuer to specify parameters such as the token name, symbol, and supply cap; (ii) $\mathsf{mint}$: used to generate tokens, typically performed by users through external tools (e.g., indexers) that coordinate inscription creation; and (iii) $\mathsf{transfer}$: used to move tokens between users, implemented through Bitcoin transactions that carry inscribed metadata indicating token movements.

We present a sample code segment illustrating the inscribed $\mathsf{Transfer}$ operation (see Listing~\ref{list-transfer} in Sec.\ref{brc20-foundation}).

\smallskip
\noindent\textbf{NOT actions, just signals!} The inscribed operations are not executable actions. Instead, they leverage the $\mathsf{OP\_RETURN}$ opcode in Bitcoin transactions, creating ``false return'' scripts that generate provably unspendable outputs. These outputs can accommodate arbitrarily formatted data written after $\mathsf{OP\_FALSE}$, ensuring that none of the inscribed information is executed on-chain. 

These inscriptions, however, serve as a signal to initiate the actual $\mathsf{transfer}$ operation via on-chain transactions.

\smallskip
\noindent\textbf{``$\mathsf{Transfer}$'' and transactions}. While the $\mathsf{deploy}$ and $\mathsf{mint}$ operations require only a single transaction (i.e., sent to oneself), the $\mathsf{Transfer}$ operation for BRC20 tokens involves two transactions (Fig.\ref{fig:transfer}): one for token inscription (\textbf{\textit{Tx1}}, sent to oneself) and another for the actual token exchange (\textbf{\textit{Tx2}}, sent to the recipient). These two transactions are executed sequentially within a very short timeframe, giving the appearance of being bundled together. Upon completion of both transactions, an off-chain indexer recalculates the final UTXO-based balance and updates the account, which corresponds to the end-user's wallet interface.

\subsection{Transfer Foundations over Bitcoin}

\noindent\textbf{UTXOs.}
Bitcoin's UTXO model serves as the backbone for tracking unspent funds in BRC20 transactions. Each UTXO is defined as a tuple, typically represented as: \(
\text{utxo} := (v, P, \mathsf{sn}),
\)
Where \( v \) denotes the value, \( P \) represents the locking conditions(encumbrance), often a public key or script, and \( \mathsf{sn} \) is the serial number ensuring uniqueness. UTXOs can be created and consumed through $\mathsf{mint}$ and $\mathsf{transfer}$ transactions. In the context of BRC20 transfers, a transaction fully consumes a set of input UTXOs and generates a new set of output UTXOs, formally expressed as:
\(
\mathsf{tx_{Transfer}} = (\vec{\text{utxo}}_{\text{inp}}, \vec{v}_{\text{out}}, \vec{P}_{\text{out}}),
\)
where \( \vec{\text{utxo}}_{\text{inp}} \) is the vector of input UTXOs, \( \vec{v}_{\text{out}} \) is the vector of output values, and \( \vec{P}_{\text{out}} \) represents the corresponding locking conditions. The UTXO model used in BRC20 ensures that tokens remain transferable while being securely tracked.

\smallskip
\noindent\textbf{Transaction fee.}
A transaction fee is determined by subtracting the total output value from the total input value. Any leftover satoshis in a transaction become the fee rewarded to miners. The higher the fee rate, the sooner the transaction will be confirmed. ransaction fees are typically measured in satoshis per virtual byte (sats/vbyte), reflecting the ratio of total fee to transaction size. To expedite delayed transactions, Bitcoin supports fee bumping mechanisms such as Replace-by-Fee (RBF)~\cite{bitcoinRBF}, which allows an unconfirmed transaction to be replaced with a higher-fee version, and Child Pays for Parent (CPFP)~\cite{bitopcpfp24}, which enables a dependent transaction to pay a higher fee to incentivize miners to confirm both parent and child together.

In the two-step transfer process of BRC20, the transaction fees associated with each step are distinct yet inherently related. Although two separate transactions are involved, users typically specify only a single fee value when initiating a transfer, which can obscure the underlying dual-transaction mechanism. The first transaction (\textbf{\textit{Tx1}}), which inscribes the intent to transfer tokens, generally requires a relatively low fee since it carries only metadata and is sent to the sender's own address. In contrast, the second transaction (\textbf{\textit{Tx2}}), which performs the actual transfer of the inscribed tokens to the recipient, tends to incur a significantly higher fee—often 3 to 5 times greater than that of \textbf{\textit{Tx1}}. Under normal conditions, this fee disparity helps ensure that \textbf{\textit{Tx2}} is confirmed promptly after \textbf{\textit{Tx1}}, maintaining the logical sequencing of the transfer process.



\smallskip
\noindent\textbf{Token balance structure.}
The BRC20 token balance structure is divided into three categories: \textit{available balance}, \textit{transferable balance}, and \textit{overall balance}, each serving a distinct purpose in managing the state of a user’s tokens~\cite{brc20experiment}.

\begin{packeditemize}
\item \textit{Available balance} ($B_{\text{avail}}$). The available balance denotes the portion of a user's tokens that is immediately liquid and can be freely spent or transferred. It reflects the confirmed token amount held by the user's address that is not involved in any pending $\mathsf{transfer}$ operations.

\item \textit{Transferable balance} ($B_{\text{trans}}$). The transferable balance represents tokens that have been inscribed with a $\mathsf{transfer}$ operation but have not yet been delivered to the intended recipient. These tokens are temporarily locked in an intermediate state and cannot be accessed, spent, or reallocated by the sender until the corresponding transfer transaction is confirmed on-chain.

\item \textit{Overall balance} ($B_{\text{total}}$). The overall balance is the sum of both the available balance and the transferable balance. It represents the total number of tokens held by the user, including tokens that are currently locked in pending transfers. The relationship between these balances can be expressed as:
\(
B_{\text{total}} = B_{\text{avail}} + B_{\text{trans}}.
\)

\end{packeditemize}

\subsection{Transfer In Action (Bitcoin Networks)} 
\label{subsec-transferNetwork}

The transfer of BRC20 over the Bitcoin network is a two-step process comprised of two transactions (as presented in Algorithm~\ref{alg:transfer1}). This design stems from the architectural constraints of Bitcoin and the nature of the inscription-based token system. Specifically, Bitcoin lacks a built-in notion of account state on Ethereum. To work around this, BRC20 transfers rely on an inscription-and-execution separation model to represent and finalize token transfers using Bitcoin’s UTXO and Taproot mechanisms.

\smallskip
\noindent\textcolor{black}{\underline{Step-\ding{172}: \textit{Inscribing the $\mathsf{transfer}$ operation.}}} The first step involves the sender inscribing a transfer request (i.e., \textbf{\textit{Tx1}}) with a transaction fee $f_{\text{tx}}$, aiming to move a specified amount of tokens $m$ to the recipient $A_r$. During this step, the sender’s available balance $B_{\text{avail}}^{A_s}$ is reduced by $m$, and an equivalent amount is added to the sender’s transferable balance $B_{\text{trans}}$.
At this stage, the tokens are placed in a pending state and are not yet executable to $A_r$. \textbf{\textit{Tx1}} alone does not affect the recipient's balances; it only logs the intention of transferring tokens. The overall balance $ B_{\text{total}} $ of the sender remains \textit{unchanged} as tokens are still under the sender’s control, albeit temporarily locked in the $B_{\text{trans}}$.

This separation arises because inscriptions are recorded as data attached to Taproot script-path spends. They can declare actions (e.g., transfer) but cannot directly move UTXOs or update on-chain token balances. Hence, while \textbf{\textit{Tx1}} inscribes the transfer intent, it does not move funds. A second transaction is required to actually spend the inscribed satoshi and complete the transfer.

\smallskip
\noindent\textcolor{black}{\underline{Step-\ding{173}: \textit{Executing the $\mathsf{transfer}$ operation.}}} In this step, the actual transfer (i.e. \textit{\textbf{Tx2}}) is executed over the network, finalizing the transfer initiated by \textit{\textbf{Tx1}}. \textit{\textbf{Tx2}} moves the specified amount $m$ from the sender’s transferable balance  $B_{\text{trans}}^{A_s}$ to the recipient’s available balance  $B_{\text{avail}}^{A_r}$. Once the \textit{\textbf{Tx2}} is confirmed, the $B_{\text{trans}}^{A_s}$ is reset to zero, and the overall balance \( B_{\text{total}}^{A_s} \) is reduced by \( m \). This marks the completion of the transfer. Simultaneously, the recipient’s \( B_{\text{avail}}^{A_r} \) and overall balance \( B_{\text{total}}^{A_r} \) are increased by \( m \). If the transfer retry limit is reached ($r >= R_{max}$) without success, the transfer is aborted to prevent retries.

\begin{algorithm}[t]
\caption{BRC20 Inscribed $\mathsf{Transfer}$ Process}

\renewcommand{\arraystretch}{1.30}
\footnotesize 
\setlength{\tabcolsep}{4pt} 

\textbf{Global Parameters:} Available balance $B_{\text{avail}}$, Transferable balance $B_{\text{trans}}$, Total balance $B_{\text{total}}$, Transaction fee $f_{\text{tx}}$. \\
\textbf{Local Parameters:} Token amount $m$, Sender's address $A_s$, Recipient's address $A_r$, Max retries $R_{\text{max}}$, Single retry $r$.

\begin{algorithmic}[1]
\State \textbf{\textcolor{black}{Step-\ding{172}: Upon initiating an $\mathsf{InscribeTransfer}$ request (\textbf{\textit{Tx1}})}}
\State Sender $A_s$ inscribes $m$ tokens for transfer to recipient $A_r$ with $f_\text{tx}$
\State Update balances for the sender's $A_s$:
    \State \quad $B_{\text{avail}}^{A_s} \gets B_{\text{avail}}^{A_s} - m$
    \State \quad $B_{\text{trans}}^{A_s} \gets B_{\text{trans}}^{A_s} + m$
    \State \quad $B_{\text{total}}^{A_s} \gets \varnothing$
    \Comment{balance unchanged}
\State Update balance for the recipient's $A_r$:
    \State \quad $B_{\text{avail}}^{A_r}\gets \varnothing$ \State \quad $B_{\text{trans}}^{A_r} \gets \varnothing$  \State \quad $B_{\text{total}}^{A_r} \gets \varnothing$ 
    \Comment{all unchanged}

\noindent\makebox[\linewidth]{\rule{0.99\linewidth}{0.3pt}} 

\State \textcolor{black}{\textbf{Step-\ding{173}: Execute transfer (\textbf{\textit{Tx2}})}} 
\State $r \gets 0$ \Comment{\textcolor{black}{initialize retry counter}}

\While{$r < R_{\text{max}}$}
    \State $A_s$ sends a transaction to $A_r$ to transfer the inscribed amount $m$
    \If{\textcolor{purple}{transaction fee} $f_{\text{tx}}$ is too low} \Comment{\textcolor{purple}{triggerring the attack}}
    \State \textbf{retry}: increase $f_{\text{tx}}$ and resend transaction
    \State $r \gets r + 1$
    \Else
        \State Update balances for $A_s$ after transfer:
            \State \quad $B_{\text{avail}}^{A_s} \gets \varnothing$ \Comment{balance unchanged}
            \State \quad $B_{\text{trans}}^{A_s} \gets 0$
            \State \quad $B_{\text{total}}^{A_s} \gets B_{\text{total}}^{A_s} - m$
        \State Update balances for $A_r$ after transfer:
            \State \quad $B_{\text{avail}}^{A_r} \gets B_{\text{avail}}^{A_r} + m$
            \State \quad $B_{\text{trans}}^{A_r} \gets 0$
            \State \quad $B_{\text{total}}^{A_r} \gets B_{\text{total}}^{A_r} + m$
        \State \textbf{break}
    \EndIf
\EndWhile
\If{$r = R_{\text{max}}$}
    \State \text{Abort}
    \Comment{failed due to over-threshold for retries}
\EndIf

\end{algorithmic}
\label{alg:transfer1}
\end{algorithm}

\section{BRC20 Pinning Attack}
\label{sec-pinningAttack}

\subsection{Transaction Pinning}\label{subsec-txpinning}

Transaction pinning~\cite{bicoinop24} refers to the case that a transaction is deliberately crafted to delay its confirmation time. 

\smallskip
\noindent\textbf{Implementing transaction pinning.}
The core principle is to intentionally keep a transaction unconfirmed in the mempool for an extended period (thereby \textit{pinning} it in place). This is achieved by manipulating transaction attributes — such as transaction fees, sequence numbers, and inputs/outputs — to control and delay the confirmation of specific transactions. 

We outline three representative techniques for implementing transaction pinning.

\begin{packeditemize}
    \item \textit{Setting low transaction fees}: In blockchain systems like Bitcoin and Ethereum, miners prioritize transactions based on the fees attached to them, processing higher-fee transactions first to maximize their rewards. By assigning a transaction fee lower than the average level, the transaction becomes less attractive to miners. As a result, the low-fee transaction remains unconfirmed for a longer duration.
    
    \item \textit{Creating non-replaceable transactions}: Transactions can be constructed to opt out of replacement protocols such as RBF) by setting the sequence number from $\mathsf{0xFFFFFFFD}$ to the maximum value ($\mathsf{0xFFFFFFFF}$). By marking a transaction as non-replaceable, the sender prevents it from being superseded by another transaction with a higher fee, maintaining its mempool position.
    
    \item \textit{Establishing dependencies}: The confirmation time can be extended by creating a chain of transactions where the inputs spend outputs from other unconfirmed transactions. If the initial transactions have low fees or are otherwise delayed, subsequent transactions cannot be confirmed until the earlier ones are processed.

\end{packeditemize}

\noindent\textbf{Pros and cons.}
Transaction pinning can serve several legitimate purposes within blockchain networks:  
(i) Users may intentionally assign low fees to non-urgent transactions in order to reduce costs;  
(ii) It can be used to prevent external parties from applying fee bumping mechanisms~\cite{bicoinop24}, retaining control over the transaction’s expected confirmation window;  
(iii) In multiparty protocols, pinning may be employed to enforce a strict transaction sequence, preventing unauthorized modifications to fees or execution order.

Transaction pinning can also be exploited to execute malicious strategies: (i) It can be used to carry out DoS attacks by flooding the mempool with pinned transactions, causing network congestion; (ii) Delayed confirmations increase the risk of front-running~\cite{daian2020flash}\cite{torres2021frontrunner}, where attackers observe pending transactions and submit their own with higher fees to ensure they are processed first.


\subsection{BRC20 Pinning Attack}

We present the BRC20 pinning attack, an attack based on the \textit{transaction pinning} and \textit{inscription transfer mechanism}.

Our BRC20 pinning attack strategically exploits a fundamental vulnerability in the protocol: the \textbf{transferable balance} state. When a token transfer is initiated via an inscription, the corresponding amount is moved from the available balance ($B_{\text{avail}}$) to the transferable balance ($B_{\text{trans}}$), where it remains in a pending state until the transfer transaction is confirmed on-chain. By applying transaction pinning techniques, an attacker can intentionally delay this confirmation, causing tokens to remain locked in $B_{\text{trans}}$ for an extended period. This creates a window of opportunity for the attacker to initiate numerous transfers and overwhelm the target’s liquidity. This denial of liquidity can halt withdrawals, disrupt operations, and create significant financial damage for the target.

Our attack proceeds in two steps, as in Algorithm~\ref{alg:low_fee_attack}.

\begin{algorithm}[t]
\caption{BRC20 Pinning Attack}

\renewcommand{\arraystretch}{1.30}
\footnotesize
\setlength{\tabcolsep}{4pt}

\textbf{Global Parameters:} $B_{\text{avail}}$, $B_{\text{trans}}$, $B_{\text{total}}$, Transaction fee $f_{\text{tx}}$, Mempool congestion $C_m$, Critical delay $T_{\text{bar}}$. \\
\textbf{Local Parameters:} Token amount $m$, Attacker's address $A_\mathcal{A}$, Target's address $A_t$, Multiple retries $R_{\text{mul}}$, Single retry $r$, Delay time $T_{\text{delay}}$.

\begin{algorithmic}[1]
\State \textbf{Step-\ding{172}: Attacker $\mathcal{A}$ initiates BRC20 token $\mathsf{transfer}$ request (\textbf{\textit{Tx}}).}
\State Attacker $\mathcal{A}$ requests to inscribe $m$ tokens to target $A_t$
\State \textcolor{purple}{Attacker $\mathcal{A}$ deliberately sets low transaction fee $f_{\text{tx}}$}
    \State \quad $f_{\text{tx}} \gets f_{\text{min}}$ \Comment{\textcolor{purple}{Minimum fee} to avoid rejection but ensure delay}

\State Update balances for the target $A_t$ normally
    
\Comment{Total balance unchanged, only internal balances shift}
    
\noindent\makebox[\linewidth]{\rule{0.99\linewidth}{0.2pt}}

\State \textbf{Check mempool congestion $C_m$.}
\If{$C_m$ is high \textbf{and} $f_{\text{tx}}$ is too low}

    \quad \tikz[baseline]{\node[anchor=base] {\textcolor{purple}{$\rightarrow$} \textit{Transaction is \textcolor{purple}{stuck} in mempool, delaying confirmation}};}
    \State Attacker $\mathcal{A}$ calculates $T_{\text{delay}}$
        \If{$T_{\text{delay}} > T_{\text{bar}}$}
            \State \textcolor{purple}{Attack succeeds}: Tokens remain in $B_{\text{trans}}^{A_t}$, locked and unusable
        \Else
            \State Go to \textbf{{Step-\ding{173}}}
        \EndIf
\Else
    \State Transaction confirmed; tokens move out of $B_{\text{trans}}^{A_t}$
    \State Balances updated normally for both $A_\mathcal{A}$ and $A_t$
\EndIf

\noindent\makebox[\linewidth]{\rule{0.99\linewidth}{0.2pt}}

\State \textbf{Step-\ding{173}: $\mathcal{A}$ repeats ($R_{\text{mul}}$) low-fee transactions to lock more tokens.}
\State Attacker continues sending multiple low-fee inscription requests to $A_t$
\For{$r = 2$ to $R_{\text{mul}}$}
    \State $\mathcal{A}$ inscribes \textcolor{purple}{$X$} tokens with \textcolor{purple}{$f_{\text{tx}}$}, where $X$ is the maximum number of $m$
    \If{transaction is \textcolor{purple}{stuck} due to high $C_m$ and low $f_{\text{tx}}$}
        
        \quad \tikz[baseline]{\node[anchor=base] {\textcolor{red}{$\rightarrow$} \textit{Target $A_t$'s tokens are \textcolor{purple}{locked} due to unconfirmed transactions}};}
        \State Update balance for $A_t$ after attack:
        \State \quad\quad\quad $\textcolor{purple}{0} \gets B_{\text{avail}}^{A_t} - X$
        \Comment{ \tikz[baseline]{\node[anchor=base] {Target $A_t$'s \textcolor{purple}{liquidity is drained}};}} 
        \State \quad $B_{\text{trans}}^{A_t} \gets B_{\text{trans}}^{A_t} + X$
        \State \quad $B_{\text{total}}^{A_t} \gets \varnothing$
        
        \State Calculate $T_{\text{delay}}$
        \If{$T_{\text{delay}} > T_{\text{bar}}$}
            \State \textbf{Exit} loop \Comment{\textcolor{purple}{Attack succeeds}, exit retry}
        \EndIf
    \Else
        \State \textbf{Next} loop; transaction confirmed
        \State \quad Balances updated normally
        \Comment{\textcolor{black}{Attack failed, next retry}}
    \EndIf
\EndFor
\State \If{max retries $r = R_{\text{mul}}$ reached without $T_{\text{delay}} > T_{\text{bar}}$}
    \State Exit
    \Comment{\textcolor{black}{Attack failed}}
\EndIf

\end{algorithmic}
\label{alg:low_fee_attack}
\end{algorithm}

\begin{packeditemize}
    \item \underline{Step-\textcolor{purple}{\ding{172}}: \textit{Inscribing \textcolor{purple}{falsified} $\mathsf{transfer}$s.}} Let \( A_t \) represent the target's address and \( A_\mathcal{A} \) the attacker's address. The attacker initiates a BRC20 token $\mathsf{transfer}$ request (\textbf{\textit{Tx}}) to \( A_t \) (\textcolor{teal}{Line 1}). Each transfer inscribes a fraction \( f \) of the target’s available balance \( B_{\text{avail}}^{A_t}(t) \) (\textcolor{teal}{Line 2}). The amount inscribed in each request \( i \) is:
    \[
    m_i = f \cdot B_{\text{avail}}^{A_t}(t), \, \text{for all } i =\{ 1, 2, \dots, N\}, f \in [0,1]
    \]
    
    At this stage, the target total's balance \( B_{\text{total}}^{A_t} \) remains unchanged, with only internal balances shifting-\( B_{\text{avail}}^{A_t} \) decreases and \( B_{\text{trans}}^{A_t} \) increases (\textcolor{teal}{Line 5}), since the inscribed operation has been confirmed. The attacker can send multiple inscription $\mathsf{transfer}$ requests (\textcolor{teal}{Line 19\&20}) to flood the network, aiming to lock up a substantial portion of the target’s liquidity.

    \item \underline{Step-\textcolor{purple}{\ding{173}}: \textit{\textcolor{purple}{Delaying} $\mathsf{transfer}$ execution.}} The attacker delays the actual transfer process (\textcolor{teal}{Line 3}) of the \textbf{\textit{Tx}} by minimizing the transaction fee \( f_{\text{tx}} \). Let \( f_{\text{tx}}^i \) represent the transaction fee for the \( i \)-th inscribed $\mathsf{transfer}$. The attacker chooses \( f_{\text{tx}} \) that is high enough to prioritize \textit{\textbf{Tx1}} but remains low enough to prevent \textit{\textbf{Tx2}} from being immediately confirmed:  
    \(
    f_{\text{tx}}^i \in [f_{\text{min}}, f_{\text{sf}}],
    \)
    where \( f_{\text{min}} \) is the threshold fee for low-priority transactions in the mempool, and \( f_{\text{sf}} \) is the upper limit at which \textit{\textbf{Tx2}} avoids quick execution (\textcolor{teal}{Line 4\&21}).

    The attacker further exploits mempool congestion, denoted as \( C_{\text{m}} \), to amplify confirmation delays (\textcolor{teal}{Line 7\&22}). Here, \( C_{\text{m}} \) is defined as the ratio between the number of unconfirmed transactions and a baseline threshold representing \textit{normal} network congestion~\cite{romiti2019deep}.

    To evaluate the effectiveness of the attack, the attacker computes the total delay \( T_{\text{delay}} \) and compares it against a pre-defined threshold \( T_{\text{bar}} \) (\textcolor{teal}{Line 9\&28}). If \( T_{\text{delay}} > T_{\text{bar}} \), the attack is deemed successful, as it results in the temporary immobilization of liquidity amounting to \( m \). In the extreme case where \( m \) equals the total available balance \( X \) of the target address \( B_{\text{avail}}^{A_t} \) (i.e., \( f = 1 \)), the entire available liquidity is effectively locked (\textcolor{teal}{Line 24}). If the delay falls short of the threshold, the attacker may retry with adjusted parameters (\textcolor{teal}{Line 20}). The attack is considered a failure only if the maximum number of attempts \( N \) is reached without achieving \( T_{\text{delay}} > T_{\text{bar}} \) (\textcolor{teal}{Line 38}).
    
\end{packeditemize}

\subsection{(Attack) Success Criteria} 

\noindent\textbf{Operational tolerance.} 
The attack is successful if the transaction with the $\mathsf{transfer}$ inscription is pinned at the first stage (i.e., \textit{inscribeTransfer}) and fails to advance to the second stage (\textit{executeTransfer}).

The key is ensuring that the delay time exceeds a certain threshold, triggering the pinned stage. Specifically, the target's tokens must remain locked in \( B_{\text{trans}}^{A_t} \) for a sufficient duration (measured as the total delay \( T_{\text{delay}} \)) to disrupt the target’s operations. The attack achieves its objective if \( T_{\text{delay}} \) surpasses the threshold \( T_{\text{bar}} \), i.e.,  $T_{\text{delay}} > T_{\text{bar}}$. 

We refer to this threshold time as \textit{operational tolerance}, which represents the minimum delay required to cause liquidity disruptions and failed withdrawals.

\noindent\textbf{Operational tolerance calculation.}
To estimate the operational tolerance \( T_{\text{bar}} \) for mainstream exchanges (CEXs, e.g., Binance, Coinbase), we consider several key factors that determine how long an exchange can endure liquidity being locked without experiencing disruptions in its operations.

\begin{packeditemize}
    \item \textit{Average transaction volume ($V$)}. The average volume of transactions that the exchange processes over a period.
    \item \textit{Required liquidity ($L_{req}$)}. The minimum amount of liquid tokens the exchange needs to meet its immediate transaction obligations.
    \item \textit{Available liquidity reserves ($L_{avail}$)}. The total amount of liquid tokens the exchange has on hand.
\end{packeditemize}

We construct a unified formula to estimate \( T_{\text{bar}} \), directly in terms of the exchange’s observed transaction volume and liquidity parameters:

\begin{equation}
    T_{\text{bar}} = \frac{L_{\text{avail}} - L_{\text{req}}}{\frac{V}{T_{\text{period}}}} = \left(L_{\text{avail}} - L_{\text{req}}\right) \cdot \frac{T_{\text{period}}}{V}
\end{equation}

We estimate that the operational tolerance window for CEXs typically falls within the range of \textbf{0.5 to 3 hours}, based on typical transaction volumes and liquidity thresholds. A detailed example for calculating $T_{\text{bar}}$ of Binance is presented in Appendix~\ref{appendix:tolerance}.

\section{Implementing Attacks Locally}
\label{sec-localExperi}


\subsection{Experiment Objectives ($\mathcal{O}$)}
We examine how the transfer amount $m$, transaction fee $f_{\text{tx}}$, mempool congestion level $C_{\text{m}}$, and number of attack attempts $N$ impact the attack effectiveness. We aim to
\begin{packeditemize}
    \item \textit{$\mathcal{O}$-1:} evaluate how an attacker can optimize the choice of $N$ in conjunction with variables $m$, $f_{\text{tx}}$, and $C_{\text{m}}$ to maximize the impact of the attack while minimizing costs;

    \item \textit{$\mathcal{O}$-2:} investigate the impact of increased fees $f_{\text{tx}}$ on the cumulative delay and liquidity disruption;
    
    \item \textit{$\mathcal{O}$-3:} examine whether a greater number of smaller transfer attempts can be as effective as fewer, larger transfers.
\end{packeditemize}

\subsection{Experiment}
\label{appendix-local}

\noindent\textbf{Configurations.}
We conducted experiments on a high-performance workstation equipped with 16 multi-core processors for parallel transaction processing, 32GB RAM for running the Bitcoin node, and a 1TB SSD for storing blockchain data. The workstation was connected to a 100 Mbps high-speed internet connection to maintain real-time synchronization with the Bitcoin testnet.

For the software setup, we deployed Bitcoin Core (v27.0) to operate Bitcoin testnet nodes and utilized the Ordinals Wallet~\cite{ordinalwallet} to facilitate BRC20 token transfers. To automate transaction generation, we developed Python scripts that leveraged libraries such as \texttt{requests} for node communication and \texttt{time} for managing transaction delays. Additionally, we used the REST API provided by Mempool.space~\cite{mempool24} to monitor real-time mempool congestion.

%

\begin{table}[h]
\centering
\caption{Experiment Parameter Settings}
\label{tab:parameter}
\renewcommand{\arraystretch}{1}
\resizebox{0.99\linewidth}{!}{
    \begin{tabular}{c|ccc|}
        \toprule
        \multicolumn{1}{c}{\textbf{Parameter}}  & \multicolumn{1}{c}{\textbf{Levels}} &  \multicolumn{1}{c}{\textbf{Range}} &  \multicolumn{1}{c}{\textbf{Results}} \\
        \midrule
        \multirow{3}{*}{\makecell{Transfer amount\\ ( $m$ )}} & \cellcolor{gray!10} Small &  \cellcolor{gray!10} $10\%$ of $B_{\text{avail}}^{A_t}$ \\
        &  \cellcolor{gray!10} Medium & \cellcolor{gray!10} $50\%$ of $B_{\text{avail}}^{A_t}$&  \multicolumn{1}{c}{\textbf{Fig.\ref{fig:pinned_tokens} }}\\
        & \cellcolor{gray!10} Large  & \cellcolor{gray!10} $100\%$ of $B_{\text{avail}}^{A_t}$ \\
        
        \cmidrule{3-4}
        
        \multirow{3}{*}{\makecell{Transaction fee \\ ( $f_{\text{tx}}$ ) }} & \cellcolor{gray!10} Minimum fee & \cellcolor{gray!10} 100 sat/vb \\
        &  \cellcolor{gray!10} Medium fee & \cellcolor{gray!10} 200 sat/vb & \multicolumn{1}{c}{\textbf{Fig.\ref{fig:cumulative_delay} }} \\
        & \cellcolor{gray!10} Higher fee &  \cellcolor{gray!10} 500 sat/vb \\
        
        \cmidrule{1-4}
        
        \multirow{3}{*}{\makecell{Mempool congestion\\ ( $C_{\text{m}}$ ) }} & \cellcolor{gray!10} Low & \cellcolor{gray!10} $25\%$ network capacity \\
        & \cellcolor{gray!10} Medium & \cellcolor{gray!10} $50\%$ network capacity & \multicolumn{1}{c}{\textbf{Fig.\ref{fig:success_rate} }} \\
        & \cellcolor{gray!10}High & \cellcolor{gray!10} $75\%$ network capacity \\
        
        \cmidrule{3-4}
        
        \multirow{3}{*}{\makecell{Number of attack attempts \\ ( $N$ ) }} & \cellcolor{gray!10} Few attempts & \cellcolor{gray!10} 2 times\\
         &  \cellcolor{gray!10} Moderate attempts & \cellcolor{gray!10}5 times & \multicolumn{1}{c}{\textbf{Fig.\ref{fig:heatmap} \& Fig.\ref{fig:violinplot}}} \\
         &  \cellcolor{gray!10} Many attempts & \cellcolor{gray!10} 10 times \\
        \bottomrule
    \end{tabular}
}

\end{table}

\smallskip
\noindent\textbf{Procedure.} Our experiments consist of three major steps.  

\begin{packeditemize}
    \item \textit{Initialization.} Set the target address $A_t$'s $B_{\text{avail}}^{A_t}$ to a constant value for consistency throughout the experiment.
    
    \item \textit{Parameter setting.} To evaluate the attack under various conditions, we generated all possible combinations of the parameters specified in Sec.~\ref{appendix-local} Table~\ref{tab:parameter}. We consider the transfer amount $m$, transaction fee $f_{\text{tx}}$, mempool congestion level $C_{\text{m}}$, and the number of attack attempts $N$, each with three levels.  
    The transfer amounts were set at 10\%, 50\%, and 100\% of the target’s available balance  $B_{\text{avail}}^{A_t}$, simulating small, medium, and large-scale attacks. Transaction fees were assigned values of 100, 200, and 500 sat/vbyte, corresponding to minimum, medium, and higher fees to assess the impact of fee manipulation on transactions. Mempool congestion levels were modeled at 25\%, 50\%, and 75\% of network capacity to represent varying network conditions from low to high congestion. The number of attack attempts was varied between 2, 5, and 10 to examine the effects of repeated attacks on the target’s liquidity. 
    By combining each level of every parameter with all levels of other parameters, we created a total of 81 unique experimental scenarios.
    
    \item \textit{Trial execution.} For each scenario, we set the specific parameter values and simulated the attack by initiating the specified number of  $\mathsf{Transfer}$ operations from the attacker’s address $A_\mathcal{A}$ to the target address $A_t$, using the designated transfer amounts $m$ and transaction fees $f_{\text{tx}}$ under the given mempool congestion level $C_{\text{m}}$. We recorded the delay time $T_{\text{delay}}$ for each transaction and observed the cumulative impact on the target’s $B_{\text{trans}}^{A_t}$ and $B_{\text{avail}}^{A_t}$, noting how much of the target’s liquidity was locked and for how long. 
    
\end{packeditemize}

\noindent\textbf{Data collection \& calculation.} We collect the following data: (i) The delay $T_{\text{delay}}$ caused by each $\mathsf{Transfer}$ operation initiated by the attacker; (ii) the total volume of tokens pinned in the target’s transferable balance $B_{\text{trans}}^{A_t}$ as a result of the $N$ attack attempts; (iii) whether each trial meets the attack’s success criteria — specifically, if $T_{\text{delay}} > T_{\text{bar}}$ and if the cumulative impact on the target’s liquidity is significant; (iv) for each parameter combination, we calculate the average liquidity outage duration and the attack success rate.

\begin{figure*}[!htbp]
    \centering
    \subfigure[Success rate]{\label{fig:heatmap}
        \includegraphics[height=4.1cm]{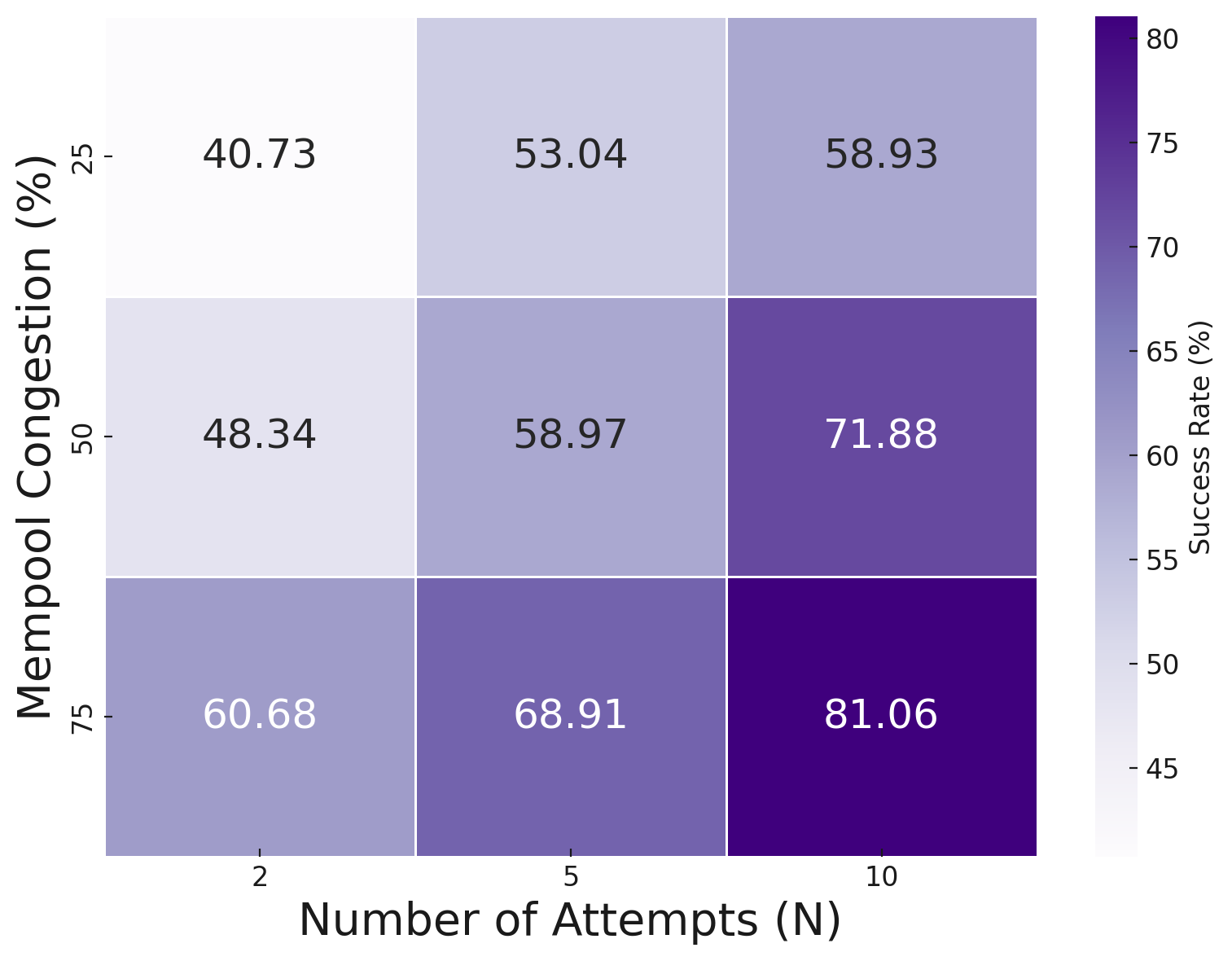}
    }
    \hspace{1cm} 
    \subfigure[Distribution of average transaction delay]{\label{fig:violinplot}
        \includegraphics[height=4.1cm]{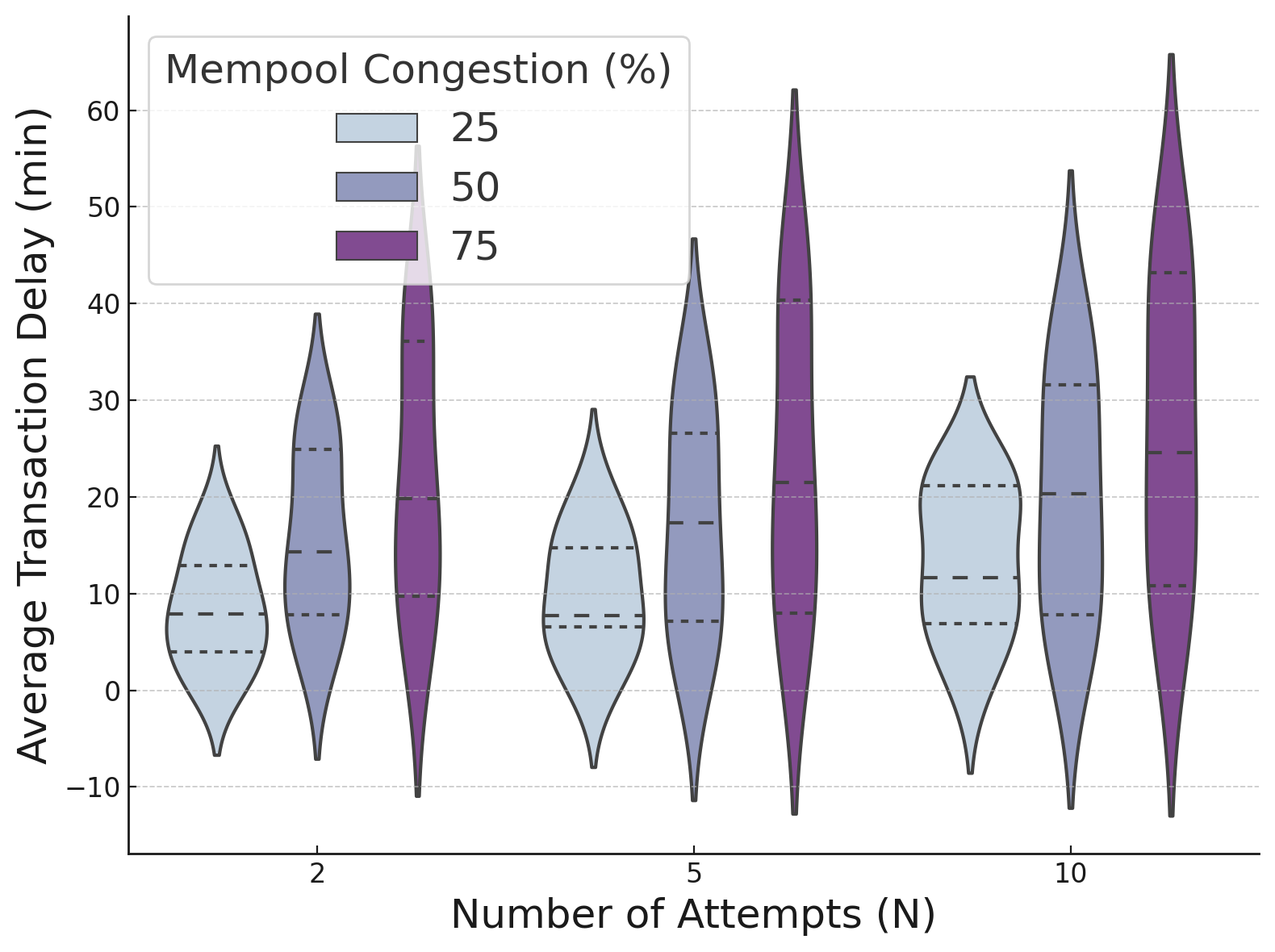}
    }
    \caption{Our evaluation for $\mathcal{O}$-1.}
\end{figure*}


\begin{figure*}[!htbp]
    \centering
    \subfigure[Transaction fee on cumulative delay]{
        \label{fig:cumulative_delay}
        \includegraphics[width=0.30\textwidth]{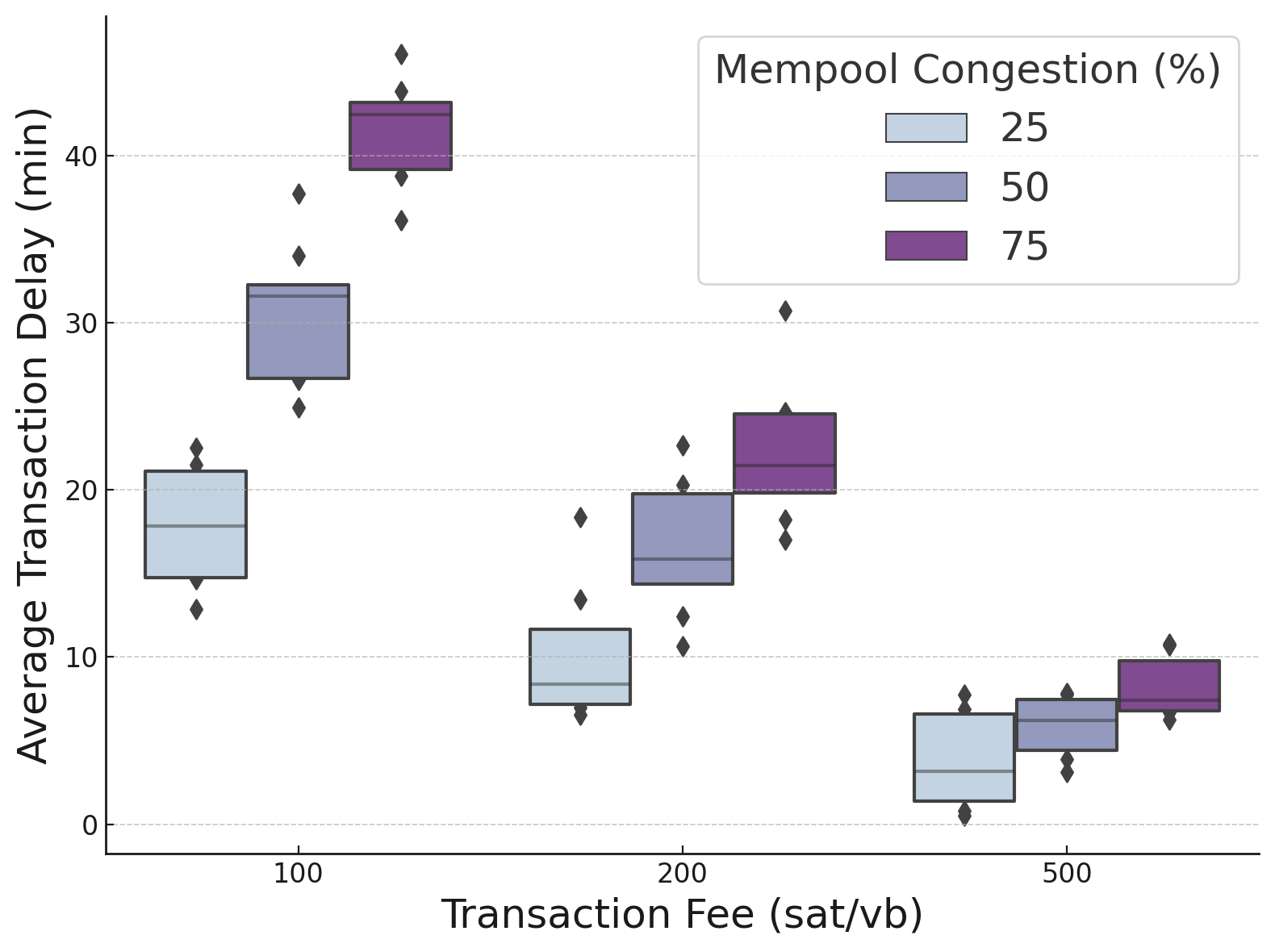}
    }
    \hspace{0.015\textwidth}
    \subfigure[Number of attempts on pinned tokens]{
        \label{fig:pinned_tokens}
        \includegraphics[width=0.30\textwidth]{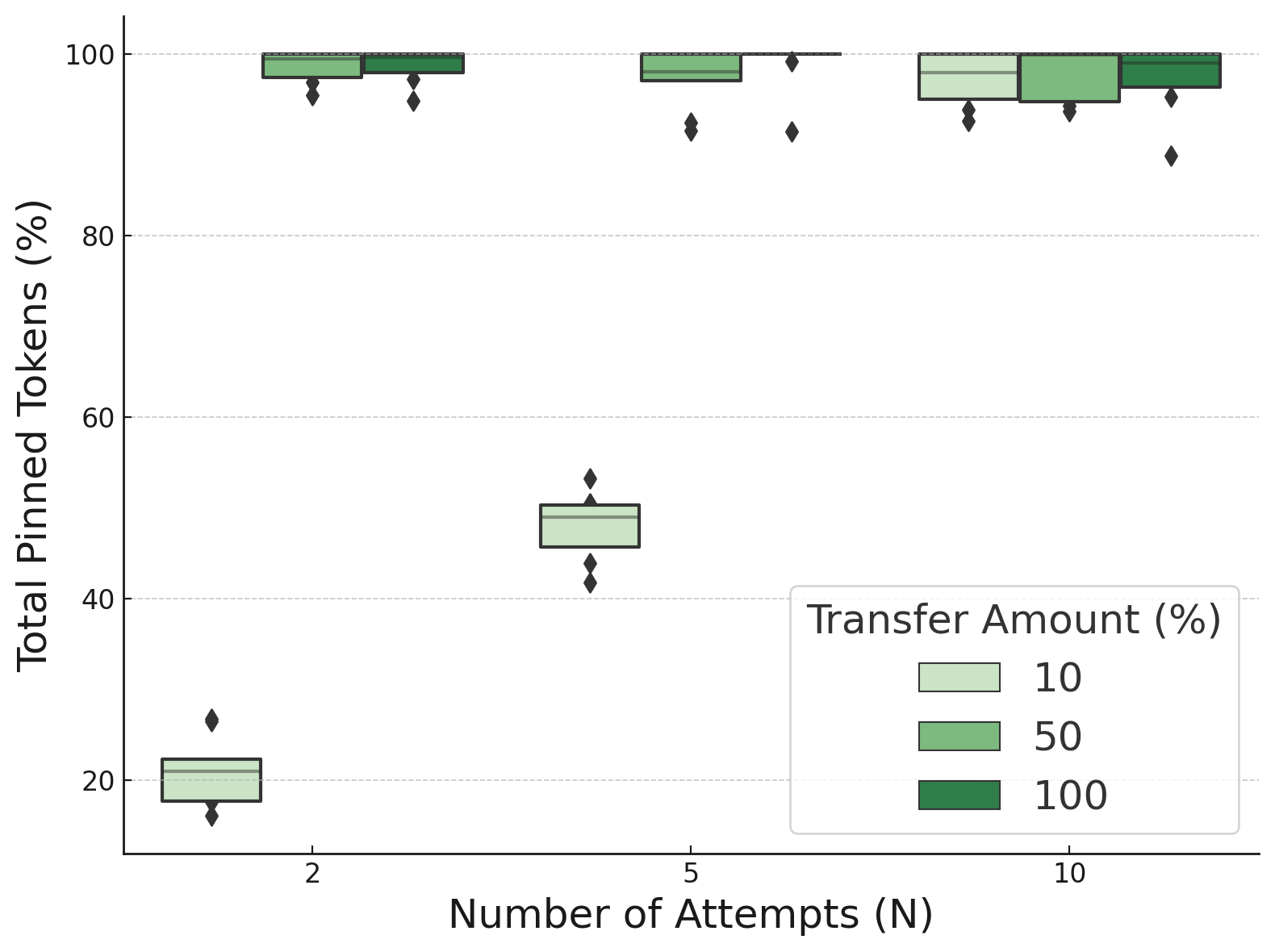}
    }
    \hspace{0.015\textwidth}
    \subfigure[Mempool congestion on success rate]{
        \label{fig:success_rate}
        \includegraphics[width=0.30\textwidth]{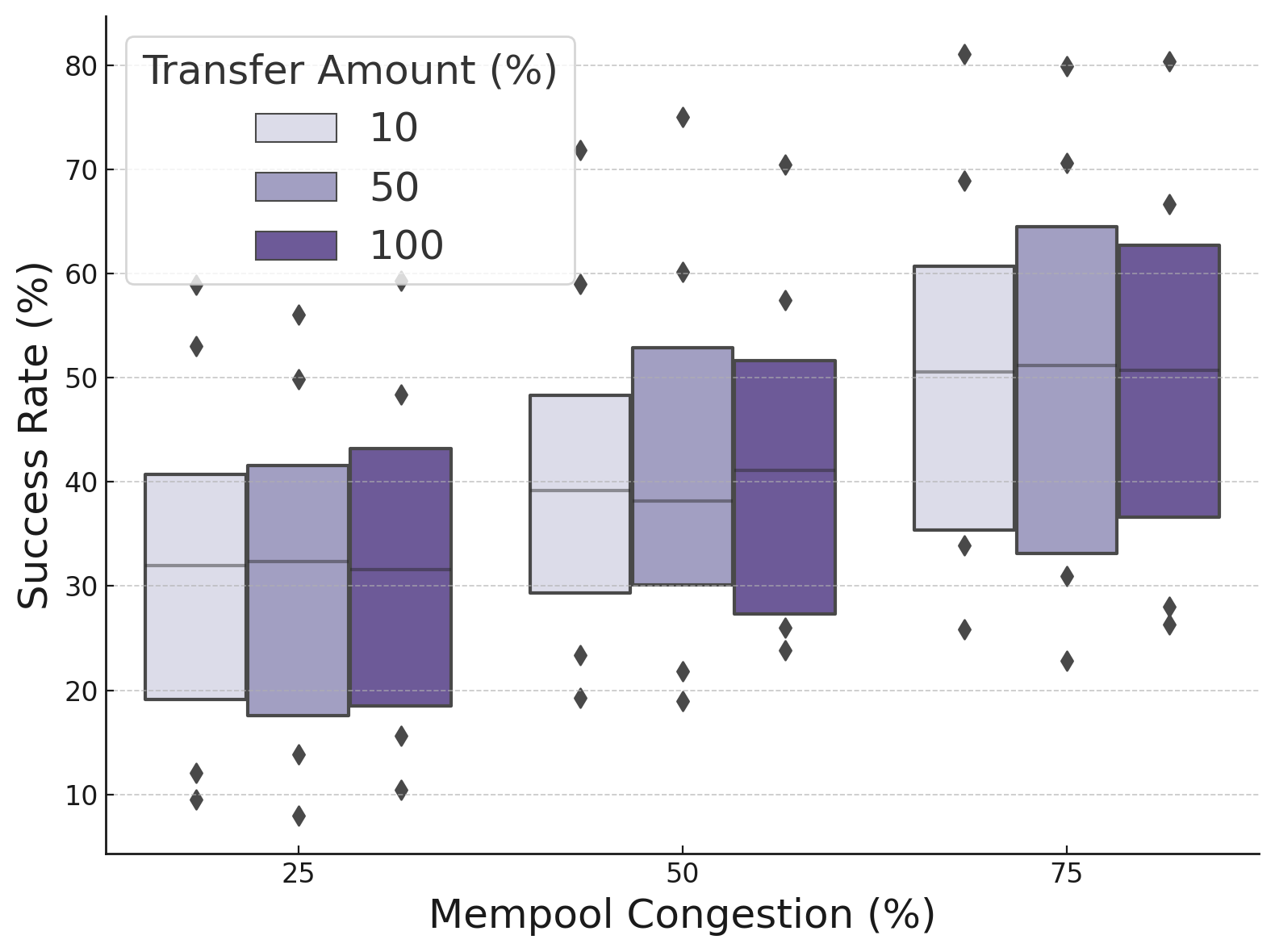}
    }
    \caption{Our evaluation for $\mathcal{O}$-2 and $\mathcal{O}$-3.}
    \label{fig:eval_figure}
\end{figure*}

\subsection{Evaluation Results}
We provide a detailed analysis of experimental data, explaining how each attack parameter impacts the attack.

\smallskip
\noindent\textbf{Evaluation on $\mathcal{O}$-1.} We first evaluate the success rate (see Fig.\ref{fig:heatmap}). We categorize success rates across three levels of mempool congestion (25\%, 50\%, and 75\%) and three scales of attack attempts (2, 5, 10 times) at a constant transfer amount and transaction fee (10\% and 100 sat/vB, respectively). 

Then, we evaluate the average transaction delay (cf. Fig.\ref{fig:violinplot}). We show the distribution of average transaction confirmation delays across different levels of mempool congestion (25\%, 50\%, and 75\%) and varying scales of attack attempts (2, 5, and 10 attempts).

\begin{packeditemize}
    \item \textit{Key observations.} (i) As $N$ increases from 2 to 10, there is a clear trend of increasing success rate across all levels of mempool congestion; (ii) higher mempool congestion levels significantly amplify the success rates of attacks. For instance, at 75\% congestion, the success rates jump from around 60.68\% at 2 attempts to an impressive 81.06\% at 10 attempts; (iii) as $N$ increases, not only does the success rate improve, but the variability in transaction delay also increases. At higher congestion levels, the delay peaks are notably higher, and the spread of delay times widens, indicating more frequent extreme values. 
    
    \item \textit{Strategic insights.} We obtain two insights: (i) the data suggests that \textbf{{maximizing the number of attempts during periods of high mempool congestion}} could be the most effective strategy for executing a pinning attack, capitalizing on the network’s vulnerabilities during high traffic to maximize disruption; (ii) while increasing $N$ at high $C_{m}$ levels provides higher success rates, it also might raise the risk of detection if not managed stealthily, requiring attackers to balance the increased success potential against the potential for heightened scrutiny.

\end{packeditemize}

\smallskip
\noindent\textbf{Evaluation on $\mathcal{O}$-2.} We evaluate (Fig.\ref{fig:cumulative_delay}) the variations in average transaction delays across different transaction fees ($f_{\text{tx}}$) under three predefined levels of mempool congestion (25\%, 50\%, and 75\%).

\begin{packeditemize}
    \item \textit{Key observations.} (i) There is a notable decrease in transaction delays as the transaction fee increases from 100 sat/vb to 500 sat/vb, demonstrating a strong inverse correlation between $f_{\text{tx}}$ and delay time; (ii) at lower fees (100 sat/vb), there is a wide range of transaction delays, particularly at higher congestion levels, indicating greater variability in transaction processing times; (iii) the reduction in delay is most pronounced when transitioning from 100 sat/vb to 200 sat/vb, suggesting that a minimal increase in fee can significantly enhance transaction processing speed.

    \item \textit{Strategic insights.} (i) Regardless of the level of congestion, \textbf{employing lower transaction fees} is a strategic approach for attackers to prolong the duration of the attack; (ii) conversely, defenders might focus on anomalous transaction patterns at lower fees during peak congestion as potential indicators of pinning or similar attacks, especially when these transactions deviate from typical user behavior in terms of speed and fee levels.
\end{packeditemize}

\smallskip
\noindent\textbf{Evaluation on $\mathcal{O}$-3.} We evaluate (cf. Fig.\ref{fig:pinned_tokens}) the impact of the number of attempts (i.e., 2, 5, 10) on the percentage of pinned tokens, segmented by the size of the transfer (small, medium, large). Additionally, we also evaluate (see Fig.\ref{fig:success_rate}) the attack success rate
 impacted by mempool congestion levels (25\%, 50\%, 75\%).

\begin{packeditemize}
    \item \textit{Key observations.} (i) An increase in pinned tokens as the number of attempts $N$ increases, with larger transfers having a more pronounced effect; (ii) the success rate is independent of the transfer size. It consistently increases with mempool congestion across all transaction sizes, reinforcing the findings from $\mathcal{O}$-1.

    \item \textit{Strategic insights.} (i) While large transfers inherently pin a high percentage of tokens in fewer attempts, \textbf{multiple small transfers} can cumulatively achieve similar disruption; (ii) under high congestion, even small transfers become effective. This is particularly useful in scenarios where avoiding attack detection is important, as transactions with large transfer amounts are more easily noticed.
\end{packeditemize}

\begin{figure*}[htbp]
    \centering
    \includegraphics[width=0.88\linewidth]{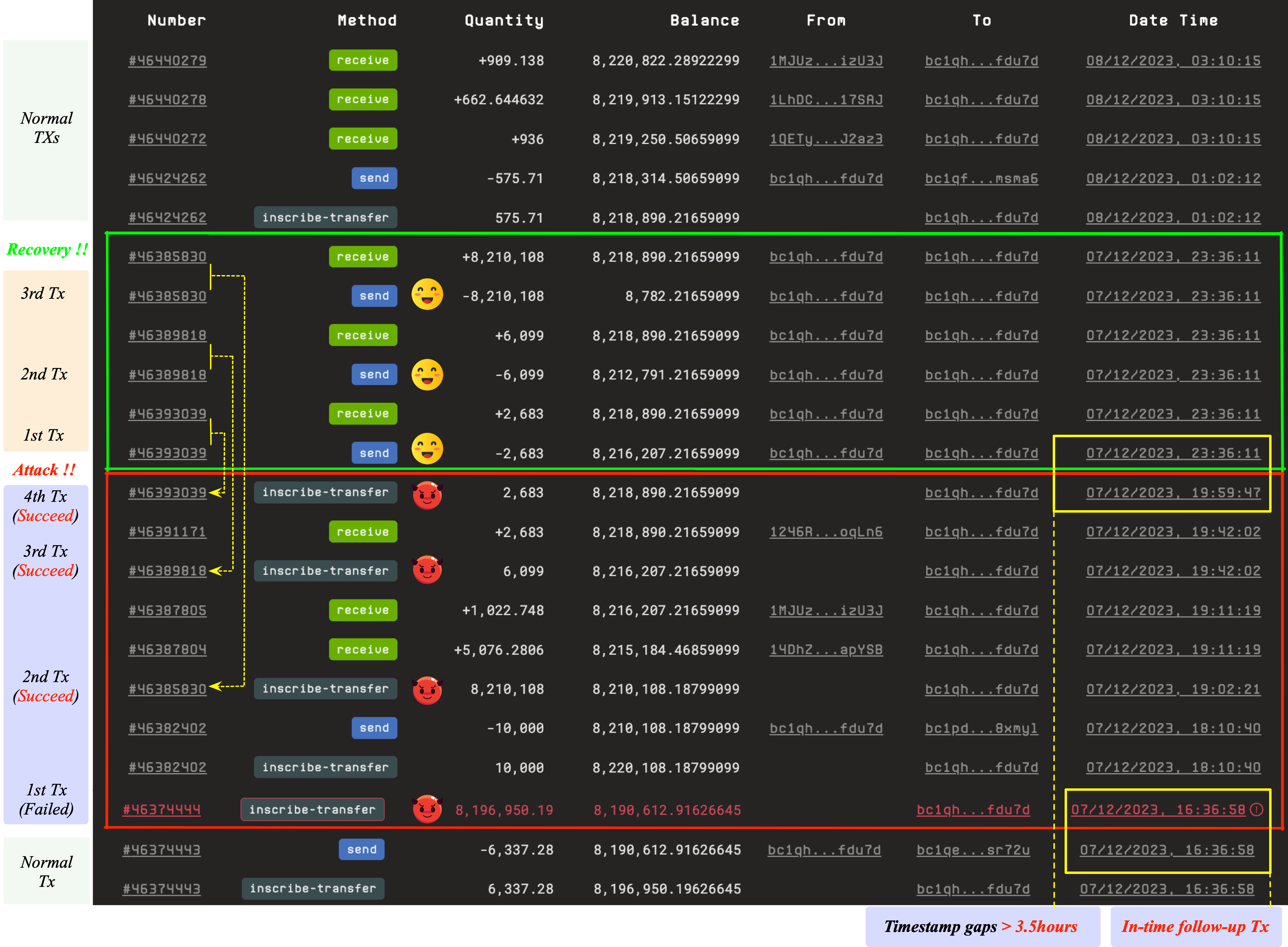}  
    \caption{
    An annotated \textbf{transaction timeline} of our pinning attack against Binance’s ORDI hot wallet, retrieved from Unisat Explorer~\cite{unisat24}. The figure (bottom to top, best viewed in color) illustrates the full transaction sequence, colour-coded into categories: \textcolor{teal!55}{normal transactions}, \textcolor{blue!65}{attack attempts}, and \textcolor{orange!90}{recovery actions}. Yellow dashed arrows indicate the logical linkage between each pair of attack and recovery transactions, matching the same amount of ORDI sent and subsequently rerouted back to the hot wallet. 
    }
    \label{fig:attack}
        \vspace{-0.18in}
\end{figure*}

\section{Coordinated Testing in Real-world}
\label{sec-attackBinance}

We present our collaborative real-world validation of the pinning attack in cooperation with Binance, using its ORDI hot wallet as the testing target. The selected wallet address\footnote{\textit{Target address: \hlhref{https://www.okx.com/web3/explorer/btc/address/bc1qhuv3dhpnm0wktasd3v0kt6e4aqfqsd0uhfdu7d}{bc1qhuv3dhpnm0wktasd3v0kt6e4aqfqsd0uhfdu7d}}} is Binance’s ORDI hot wallet, which was jointly identified as an appropriate test target.

\smallskip
\noindent\textbf{Reasons for selecting ORDI.}
ORDI ranks as the \textbf{No.1}\footnote{As of the time of writing, ORDI’s MarketCap is \href{https://coinmarketcap.com/view/brc-20/}{\$711M}.} BRC20 token in terms of market capitalization and on-chain activity. Binance has the \textbf{largest } trading volume for ORDI, occupying 26.24\%\footnote{Data from CoinMarketCap: \href{https://coinmarketcap.com/currencies/ordi/}{https://coinmarketcap.com/currencies/ordi/}.}\setcounter{myfootnote}{\value{footnote}} of the total market volume, making its ORDI wallet the most active. The second-largest volume is held by OKX, occupying 15.99\%\footnotemark[\value{myfootnote}].

\smallskip
\noindent\textbf{Reasons for selecting a targeted address.}
We selected this address as our primary target for attacking for several reasons: 
(i) At the time of testing, this selected wallet held \underline{8,196,950} ORDI tokens (valued at approximately US\$\underline{9,052,645}), making it representative for evaluating the real-world implications of a pinning attack. (ii) OKX explorer~\cite{okx24} indicates that it is one of the most active addresses for ORDI, handling deposits and withdrawals for a large user base. By focusing on an address with high transactional throughput (150 txs/day\footnote{Data calculated by $\mathtt{Total\_txns}/\mathtt{Total\_days}$, \href{https://www.okx.com/web3/explorer/btc/address/bc1qhuv3dhpnm0wktasd3v0kt6e4aqfqsd0uhfdu7d}{OKX}}), we can easily assess how the attack disrupts normal operations.

\smallskip
\noindent\textbf{Attack tools.} 
We used the Ordinals wallet interface~\cite{ordinalwallet} to interact with the ORDI protocol, creating multiple $\mathsf{inscribeTransfer}$ transactions targeting the wallet.
We employed the Mempool.space API~\cite{mempool24} to monitor Bitcoin’s mempool congestion in real-time.
We also used the Ordiscan explorer~\cite{ordiscan24} to verify the status of the ORDI wallet during the attack and Unisat~\cite{unisat} to track specific transaction details and evaluate the impact on Binance’s operations.




\subsection{Execution of Attack}\label{subsec-exeAttack}

\noindent\textbf{Attack overview.} In collaboration with Binance’s research team, we conducted a series of controlled and monitored tests on the ORDI hot wallet to validate the effectiveness of the pinning attack under real-world conditions (Fig.\ref{fig:attack}). These tests involved constructing transactions with deliberately low fees and large token amounts under periods of high mempool congestion, with the aim of evaluating whether the transferable balance could be immobilized.

During the coordinated execution, we successfully demonstrated that the wallet’s liquidity could be temporarily locked, simulating an operational disruption in ORDI withdrawals. The controlled test resulted in a transient suspension of ORDI withdrawal functionality for approximately \textbf{3.5 hours}, during which the wallet's transferable balance was fully occupied. The experiment was closely monitored by both parties, and upon completion, Binance promptly executed recovery transactions to restore normal operations.

We visualise the test process in Fig.\ref{fig:attack}. Transactions used in the controlled validation are marked in \textcolor{purple}{purple}, recovery operations are in \textcolor{teal}{teal}, and regular user-driven inscriptions are shown in black.

\smallskip
\noindent\textbf{First attempt} (\textcolor{black}{\textbf{\textit{1st Tx}}} in Fig.\ref{fig:attack}).
The first attack (\textcolor{black}{\textbf{\textit{1st Tx}}}, $\mathsf{inscription ID}$: \hlhref{https://ordiscan.com/inscription/46374444}{\textcolor{purple}{\#46374444}}) took place on Dec. 7, 2023, at 16:36:58, targeting Binance’s ORDI wallet to inscribe a transfer of 8,196,950 ORDI tokens, the entire balance of the wallet. Given the high mempool congestion (149.48\% capacity), we executed the attack using a low-fee and large-amount strategy (200 sats/vB, approximately 0.000278 BTC), significantly lower than the average transaction fee of 0.00051525 BTC~\cite{blockcom24}, to delay transaction confirmation. We provided more detailed attack information in Appendix~\ref{app:attack-details}.


The first attack attempt was unsuccessful due to an unexpected concurrent transaction. While we initiated the attack to inscribe a transfer of 8,196,950 ORDI tokens, a withdrawal transaction ($\mathsf{Inscription ID}$: \href{https://ordiscan.com/inscription/46374443}{\textcolor{black}{\#46374443}}) was confirmed nearly simultaneously, reducing the Binance wallet’s balance by 6,337 ORDI tokens.

This reduction in the wallet’s balance resulted in our attack attempting to lock more tokens than were available in the wallet — specifically, our inscribed amount exceeded the wallet’s remaining total by the amount withdrawn. Since the intended amount to be locked could not surpass the wallet's actual holdings, the attack was ultimately unsuccessful.

\smallskip
\noindent\textbf{Subsequent attempts} (\textcolor{black}{\textit{\textbf{2nd-4th Txs}}} in Fig.\ref{fig:attack}).
After the initial failure, we proceeded to launch three successive attacks approximately one hour later\footnote{During this interval, a normal ORDI transfer (\href{https://ordiscan.com/inscription/46382402}{\textcolor{black}{\#46382402}}) occurred.}. We continued to use an attack strategy with large inscription transfer amounts and low fees. Inscriptions with higher amounts effectively maximize liquidity locking, while low fees prolong the transaction's presence in the mempool.

At 19:02:21, we initiated our second attack (\textcolor{black}{\textbf{\textit{2nd Tx}}}, $\mathsf{inscription ID}$: \hlhref{https://ordiscan.com/inscription/46385830}{\textcolor{purple}{\#46385830}}) using a fee rate of 201 sats/vB and successfully locked 8,210,108 ORDI tokens, nearly the entire balance of the target address. This rendered the withdrawal functionality of the target wallet \textbf{inoperative}, as all tokens were locked in $B_{\text{trans}}^{A_t}$.

\begin{table}[h]
\centering
\renewcommand{\arraystretch}{0.95}
\resizebox{\linewidth}{!}{
    \begin{tabular}{|ll}
    \textbf{Attack}  & \textcolor{black}{\textbf{\textit{2nd Tx}}} \\
    Block ID & \#820,119 \\
    Inscription ID & \hlhref{https://ordiscan.com/inscription/46385830}{\textcolor{purple}{\#46385830}} \\
    Transaction ID & de31d83400784112a3e...7f14087b351971f0e \\
    Source Address &  bc1pclw7e5azt70plf...xgesc4jh7s9vta4y (Taproot address)  \\
    Target Address & bc1qhuv3dhpnm0...sd0uhfdu7d (Binance's ORDI address) \\ 
    Amount & Input: 0.00028146 BTC, Output: 0.00000546 BTC  \\
    Fee rate: & 201 sats/vB \\
    Fee used: & 27,800 sats \\
    Attack result & \textcolor{purple}{Success} \\
    \end{tabular}
}
\vspace{-0.2in}
\end{table}

Subsequently, at 19:11:19, the target address received two consecutive token transfers totalling 5,076 and 1,022 tokens ($\mathsf{inscription IDs}$: \href{https://ordiscan.com/inscription/46387804}{\textcolor{black}{\#46387804}} and \href{https://ordiscan.com/inscription/46387805}{\textcolor{black}{\#46387805}}). These two transactions appeared to be legitimate inscriptions initiated by Binance. OKLink\footnote{OKLink is a blockchain explorer that provides \underline{address associations} for transactions; see \href{https://www.oklink.com/btc/tx/309d7139cfd08de502e99e99456e2d6333f8260d2f4a72831c709622d0b498a9}{\textcolor{black}{\#46387804}} and \href{https://www.oklink.com/btc/tx/ebe891e8e5cd4c954dedf3b0395c6929fc06c250bd24da734e304a0458743f45}{\textcolor{black}{\#46387805}}, respectively.} indicates that the sending addresses are associated with Binance’s internal wallets. The purpose of these transfers was to replenish the target wallet’s available balance $B_{\text{avail}}^{A_t}$ by transferring extra ORDI tokens to it, thereby restoring withdrawal functionality for users.

At 19:42:02, we launched the third attack (\textcolor{black}{\textbf{\textit{3rd Tx}}}, $\mathsf{inscription ID}$: \hlhref{https://ordiscan.com/inscription/46389818}{\textcolor{purple}{\#46389818}}), targeting exactly the sum of the newly received 6,099 tokens, which once again locked all liquidity at the target address. 


Concurrently with this third attack, the target address received an additional transfer of 2,683 tokens ($\mathsf{inscription ID}$: \href{https://ordiscan.com/inscription/46391171}{\textcolor{black}{\#46391171}}), prompting us to initiate one more attack (\textcolor{black}{\textbf{\textit{4th Tx}}}, $\mathsf{inscription ID}$: \hlhref{https://ordiscan.com/inscription/46393039}{\textcolor{purple}{\#46393039}}) at 19:59:47, with an equal amount of 2,683 tokens. 

\smallskip
The above coordinated tests demonstrated that it was possible to lock the transferable balance of the target address, thereby preventing standard operational transactions such as withdrawals and receipts during the testing period. For around 3.5 hours following the initiation of the test sequence, no outgoing transactions were observed at the target address. We witnessed that those repeated attempts were able to drain out wallet-level liquidity. The outcome validated the practical feasibility of our pinning attack. 

\smallskip
\noindent\textbf{Liquidity recovery.}  
As part of the coordinated testing process, we (with Binance) conducted instant and controllable recovery actions to verify that liquidity could be restored after transferable balance had been pinned.

In the two-step inscription-based transfer model, the initial transaction (\textbf{\textit{Tx1}}) only inscribes the intent to transfer, while the actual transfer is executed in a subsequent transaction (\textbf{\textit{Tx2}}). If \textbf{\textit{Tx2}} remains unconfirmed, the tokens reside in a non-finalized, transferable state under the sender’s control. To recover the liquidity, Binance initiated new \textbf{\textit{Tx2}} transactions that corresponded to the pinned \textbf{\textit{Tx1}} inscriptions, thereby rerouting the tokens back to its own wallet address.

These recovery operations were successfully executed and are visible in Fig.\ref{fig:attack}. The recovery transactions are highlighted in \textcolor{teal}{teal}. Specifically, three recovery transactions (confirmed in $\mathsf{Block}$ \hlhref{https://mempool.space/block/0000000000000000000366d3fc3865713bec74c93b4e56cf71dcf95c0c72d473}{\#820,133},  with inscription IDs \hlhref{https://ordiscan.com/inscription/46393039}{\textcolor{teal}{\#46393039}}, \hlhref{https://ordiscan.com/inscription/46389818}{\textcolor{teal}{\#46389818}}, and \hlhref{https://ordiscan.com/inscription/46389830}{\textcolor{teal}{\#46389830}}) transferred a total of 8,218,890 ORDI tokens from the transferable balance back into the available balance. Normal transactional activity, including deposits and withdrawals, resumed shortly after the completion of this recovery phase, as evidenced by follow-up transactions (i.e., $\mathsf{Inscription\, ID}$: \href{https://ordiscan.com/inscription/46389830}{\textcolor{black}{\#46424262}}).


\section{Attack Insights}
\label{sec-discussion}


\subsection{Potential Defenses}

\noindent\textbf{Defense strategies.} We propose and empirically evaluate several practical defense mechanisms to mitigate the risk of BRC20 pinning attacks. These strategies were tested in a controlled environment replicating realistic mempool congestion and transaction dynamics.

\begin{packeditemize}
    \item \textit{Abnormal transaction monitoring.} We implemented a lightweight anomaly detection module that monitors live transaction streams and flags transfer requests exhibiting atypical fee-to-value ratios or repetitive low-fee patterns. In our testbed simulating exchange wallet behavior under attack, this mechanism successfully detected all simulated pinning attempts with a 98.7\% accuracy rate. Once flagged, such transactions can be either temporarily suspended or queued for manual verification, helping to prevent large-scale liquidity locks before confirmation.
    
    \item \textit{Dynamic fee adjustments.} We simulated a mempool-aware dynamic fee policy where the exchange adjusts the minimum acceptable fee threshold in real time based on mempool congestion ($C_m$). Our evaluation showed that this mechanism reduced average pinning duration by 63.2\%, while making repeated low-fee attacks economically unviable. This ensures that legitimate transactions are prioritized by miners, minimizing attack effectiveness.
    
\end{packeditemize}

\smallskip
\noindent\textbf{\textcolor{red}{\dangersign}} \noindent\textbf{Limitation in defense.} We have to admit that, despite the promising results of these mechanisms, certain protocol-level constraints inherent to BRC20 limit the scope of defense. Specifically, the BRC20 design allows for $\mathsf{transfer}$ inscriptions to be broadcast and remain unconfirmed, which is a legitimate and intended behavior. As a result, it is fundamentally difficult to distinguish malicious pinning from benign pending transactions solely based on mempool state.

Moreover, BRC20 lacks built-in programmability and intent-awareness. All transfer inscriptions are treated identically, without additional metadata, signatures, or logic to differentiate valid user transfers from manipulative operations. The absence of programmability in Bitcoin prevents integration of advanced contract-level mitigations such as time locks, multisigs, or conditional execution.

\subsection{Attack Applicability}\label{subsec-applicable}

\noindent\textbf{Applicability to 90\% BRC20 protocols} (deferred Table~\ref{tab:brc-tokens}).
Our attack is inherently applicable to other BRC20 tokens (\textit{upper}  Table~\ref{tab:brc-tokens}) due to the uniformity of their two-step transactional mechanism: the first transaction records the intent to transfer tokens by embedding data into a Bitcoin transaction (via $\mathsf{opcode}$), and the second actualises the transfer when the transaction is confirmed. The two transactions appear to be bundled, but in fact, they are \textbf{separate} from each other. Our attack pins the first transaction, obstructing the execution of the second one, where actual token exchanges take place. 

As an early warning, we need to highlight that the vulnerabilities exploited in this work are not specific to a single token but are systemic, affecting a wide range of tokens, including but not limited to those listed in the table.

\smallskip
\noindent\textbf{Inapplicability to any inscription protocols?}
Our attack does not apply to EVM-based inscriptions (\textit{bottom} Table~\ref{tab:brc-tokens}), such as Ethscriptions on Ethereum~\cite{ethscription24}, due to fundamental differences in transaction structures and execution logic.

Ethscriptions embed data directly into the $\mathsf{calldata}$ field of transactions~\cite{xiong2025talking}. For example, a Base64-encoded image is converted to a data URI, then into a hexadecimal string, and included in a native ETH transfer’s $\mathsf{calldata}$. The recipient address becomes the owner of the inscribed token.

More broadly, EVM-compatible blockchains use a \textbf{single-}transaction model for token transfers: \textit{inscription and execution occur atomically within the same transaction}. This eliminates the deferred execution step exploited by our BRC20 attack, making the same vulnerability unexploitable.




\begin{table*}[t]
\centering
\caption{\textbf{Malicious mempool-layer exploits:}
Unlike prior works that focus primarily on frontrunning, fee replacement, or gas bidding to reorder transactions,  our attack uniquely targets the dynamic transfer mechanism and exploits its intermediate \textit{transferable} state. By crafting a low-fee Tx that remains in the mempool, we are able to lock the UTXO and disrupt entire liquidity.}

\label{tab:attack_taxonomy}

\renewcommand{\arraystretch}{1.1}
\resizebox{\textwidth}{!}{%
\begin{tabular}{c|cc|cc|c}

\toprule

\multicolumn{1}{c}{\textbf{Attack Type}} & 
\multicolumn{1}{c}{\textbf{Target}}  & 
\multicolumn{1}{c}{\textbf{Goal}} &
\multicolumn{1}{c}{\textbf{Exploit Vector}} & 
\multicolumn{1}{c}{\textbf{Tactics}}  & 
\multicolumn{1}{c}{\textbf{Applicable Platform}}   
\\ 

\midrule
\cellcolor{magenta!19} \textbf{Our pinning attack} & \cellcolor{magenta!19}  BRC20 Inscriptions & \cellcolor{magenta!19} Lock token state & \cellcolor{magenta!19} Two-step transfer flow & \cellcolor{magenta!19} Withdrawal pinning via fee control &\cellcolor{magenta!19}  Bitcoin (UTXO) \\ 

\midrule

Replace-by-Fee~\cite{li2023transaction,txpinHTLC} & Low-fee txs & Prioritize adversary tx & Replaceable fee logic & Strategic re-broadcasting & Bitcoin (UTXO) \\ 

Tx malleability~\cite{decker2014bitcoin,andrychowicz2015malleability} & Transaction ID & Break tx linkage & Sig alteration pre-mining & Modify tx hash in-flight & Bitcoin (UTXO) \\

PSBT pinning~\cite{qi2025brc20} & Partially signed PSBTs & Block token flow & Metadata misuse & PSBT hold + RBF denial & Bitcoin (UTXO) \\

\midrule

Frontrunning~\cite{daian2020flash,zhou2021high,zhang2022frontrunning} & Pending txs & Insert attacker tx & Fee auction & Backrun via gas price race & EVM-compatible \\ 

MEV arbitrage~\cite{eskandari2020sok,heimbach2022sok,yang2024sok} & Tx order in block & Maximize extractor profit & Miner collusion & Ordering or censorship & EVM-compatible \\

Sniping bots~\cite{cernera2023token,cernera2023ready,cernera2024warfare} & NFT/token mints & Capture rare assets & Gas wars & Auto gas inflation & EVM-compatible \\  

Imitation attacks~\cite{qin2023blockchain} & Deployed contracts & Copy logic for gain & Interface mirroring & Deploy-copy-exploit cycle & EVM-compatible \\ 

\midrule

Timejacking~\cite{zhang2023time} & Node timestamps & Skew block time & Fake timestamps & Validator clock drift & All blockchains \\

Double-spending~\cite{negy2020selfish,feng2019selfish,eyal2018majority} & UTXO or state & Spend same asset twice & Branching or tx race & Conflicting tx broadcast & All blockchains \\

Mempool DDoS~\cite{wu2020survive,saad2019shocking,saad2019mempool} & Mempool queue & Starve valid txs & Spam floods & Saturation via low-priority txs & All blockchains \\

\bottomrule

\end{tabular}%
}
\vspace{-0.4em}
\end{table*}

\subsection{Attack Bounds} 
\noindent\textbf{Transaction fee.}
The fee bounds in our attack are governed by two dynamic thresholds: the minimum fee \( f_{\text{min}} \) and the safety fee \( f_{\text{sf}} \). The lower bound \( f_{\text{min}} \) ensures that \textbf{\textit{Tx1}} remains in the mempool without being dropped, typically ranging from 1 to 10 sat/vB under low congestion. Setting a fee slightly above this threshold keeps \textbf{\textit{Tx1}} active while delaying the confirmation of \textbf{\textit{Tx2}}. The upper bound \( f_{\text{sf}} \), often 2 to 2.5 times \( f_{\text{min}} \), is the highest fee below which miners are unlikely to prioritise the transaction. Both thresholds are influenced by real-time network congestion and platform-specific policies. Wallets like Ordiscan and OKX may enforce differing minimum fee rules.



\smallskip
\noindent\textbf{Suspension tolerance.}
It is important to acknowledge that suspension tolerance varies across platforms, network conditions, and external factors. Our estimation is derived from a large-scale dataset covering multiple time periods and congestion levels on OKX, providing a generalised reference window. However, platform-specific thresholds (such as those of Binance, Ordiscan, and OKX) can differ significantly due to variations in transaction volume, liquidity management policies, and user activity. Moreover, external events like network upgrades or sudden spikes in token demand may temporarily alter these tolerances, thereby affecting the timing of pinning-based attacks.

\subsection{Open Problems and ``To Be Explored''}

\noindent\textbf{Pinning attack extension.} 
Our findings suggest that platforms leveraging PSBT (Partially Signed Bitcoin Transactions~\cite{bip174}), such as Ordinals Wallet, may also be vulnerable to pinning attacks. Since PSBTs remain in a partially signed state until fully confirmed, attackers can exploit low-fee pinning to monopolise pending transactions, preventing others from finalising purchases. This highlights the need for new timeout or auction mechanisms to prevent indefinite PSBT lockups and ensure fair market access.

\smallskip
\noindent\textbf{Fee distribution in transfers.} An open challenge lies in understanding how transaction fees ($f_\text{tx}$) are allocated between the two steps of BRC20 token transfers: \textbf{\textit{Tx1}} (which inscribes the transfer intent) and \textbf{\textit{Tx2}} (which executes the actual transfer). Our empirical observations suggest that \textbf{\textit{Tx2}} typically incurs a fee 3 to 5 times higher than \textbf{\textit{Tx1}}. However, the underlying protocol logic governing this fee allocation remains poorly documented. Clarifying this mechanism is critical for designing effective defenses, as adversaries can exploit the fee asymmetry to selectively delay \textbf{\textit{Tx2}} and launch pinning attacks.

\subsection{Related Work (Attack-focused)}
\label{sec-rw2}

\noindent\textbf{Bitcoin pinning attack.} The attack is a strategy that deliberately keeps an unconfirmed transaction in the mempool and delays its confirmation on-chain~\cite{txpin24}. This introduces vulnerabilities for services that depend on timely transaction processing. For example, (i) in payment channels like the Lightning Network~\cite{txpinHTLC}, participants often need to close the channel within a specified timeframe, and an attacker can pin transactions to prevent the channel from closing on time; (ii) in escrow services, where time-based conditions dictate when funds are released, an attacker can exploit these conditions by delaying the confirmation of critical transactions. Our attack follows a similar principle, wherein an attacker can pin an unconfirmed transaction ($\mathsf{inscribeTransfer}$) to prevent it from advancing to the settlement stage ($\mathsf{Transfer}$). 


\smallskip
\noindent\textbf{Mempool-based attacks} (Table~\ref{tab:attack_taxonomy}).
Mempool-layer attacks have been extensively studied across both account-based and UTXO-based blockchains. On Bitcoin and other UTXO chains, RBF-based exploits~\cite{li2023transaction,txpinHTLC} enable adversaries to overwrite pending transactions via fee replacement, and transaction malleability~\cite{decker2014bitcoin,andrychowicz2015malleability} disrupts transaction linkage by altering non-finalized transaction IDs. Recent work has also revealed vulnerabilities in the PSBT layer~\cite{qi2025brc20}, where metadata injection and RBF logic can be combined to hijack partially signed inscriptions. In EVM-compatible systems, frontrunning~\cite{daian2020flash,zhou2021high,zhang2022frontrunning} and MEV arbitrage~\cite{eskandari2020sok,heimbach2022sok,yang2024sok} exploit transaction ordering and fee dynamics to gain unfair advantage, often leveraging miner-controlled sequencing or backrunning techniques. Sniping bots~\cite{cernera2023token,cernera2023ready,cernera2024warfare} automate fee escalation to capture NFT/token mints, while imitation attacks~\cite{qin2023blockchain} clone contract interfaces to deceive users and intercept transactions. Timejacking~\cite{zhang2023time} and double-spending~\cite{negy2020selfish,feng2019selfish,eyal2018majority} further illustrate how timestamp or consensus-level manipulations affect transaction validity, while DDoS-style attacks~\cite{wu2020survive,saad2019shocking,saad2019mempool} leverage mempool flooding to degrade network availability.

\smallskip
\noindent\textbf{Token incidents.}  Token incidents are downstream ``victims'' of broader system vulnerabilities (mempool-layer included) and manifest in various forms, including design flaws in intentional arbitrage (e.g., flashloan~\cite{chen2024flashsyn}, MEV exploits~\cite{heimbach2022sok}), complex strategies (primarily in DeFi~\cite{zhou2023sok}), malicious fraud (e.g., Ponzi schemes~\cite{chen2021sadponzi}, scams~\cite{xia2021trade}, rug pulls~\cite{cernera2023token}), as well as the unpredictable collapse of major players (e.g., SVB~\cite{wang2024cryptocurrency}, FTX~\cite{fu2023ftx}). Most of these incidents occur with ERC20 tokens\footnote{We majorly focused on upper-layer tokens rather than native platform tokens, largely due to the more complex attack strategies involved.} in the Ethereum ecosystem. Our attack reports the first upper-layer token incident within Bitcoin.

\label{sec-rw}

    
    

\section{Conclusion}
\label{sec-conclusion}

We present the \textit{BRC20 pinning attack}, the first practical demonstration of a vulnerability in the BRC20 token transfer mechanism. This flaw arises from the UTXO-based inscription model, where two loosely coupled transactions (\textit{\textbf{Tx1}} and \textit{\textbf{Tx2}}) are required to complete a transfer. Under specific fee configurations and mempool congestion, the second transaction (\textit{\textbf{Tx2}}) can remain unconfirmed, leaving tokens in a frozen state and effectively locking liquidity.

We performed local evaluations followed by coordinated real-world testing in collaboration with Binance. We focused on the most active BRC20 hot wallet (i.e., ORDI). Our tests confirmed that liquidity can be locked and withdrawals halted for up to 3.5 hours. Our broader analysis shows that over \textbf{90\%} of BRC20 tokens are similarly vulnerable due to shared inscription logic. We also proposed several early-stage mitigation solutions. We encourage further engagement in developing stronger safeguards for BRC20s.




\bibliographystyle{unsrt}
\bibliography{bib.bib}

\appendix

\subsection{Extended Related Work}

\noindent\textbf{Existing studies on BRC20.} Wang et al. \cite{wang2023understanding} analyze the activities surrounding BRC20 hype and conduct a multi-dimensional evaluation of market function and social sentiment. Louis~\cite{bertucci2024bitcoin} investigates the transaction fee fluctuation of Bitcoin Ordinals and assesses its impact on the Bitcoin network. Kiraz et al.~\cite{kiraz2023nft} introduce a proof system leveraging recursive SNARKs to establish the linkage of two Bitcoin transactions. While they do not explicitly mention BRC20 or token inscription, the method for transferring NFTs via Bitcoin adheres to the same underlying principle. Wen et al.~\cite{wen2024modular} designed a decentralized Bitcoin indexer, a type of distributed relays. They leverage Verkle trees~\cite{kuszmaul2019verkle} to prove the states, replacing committee indexers with heavy computations with lightweight users.  Wang et al.~\cite{wang2024bridging} proposed a unilateral bridge to allow initiating operational actions from the Bitcoin side to Ethereum smart contracts.  Additionally, several studies have initiated discussions on BRC20 within the broader scope of NFTs, including analyses of NFTs themselves~\cite{razi2023non,allen2023web3,conlon2023problem}, price fluctuations~\cite{szydlo2024characteristics}, and market sentiment~\cite{baklanova2023investor}.

\smallskip
\noindent\textbf{Further studies stemming from BRC20.} Several technical routes have emerged: (i) A line of protocols modify the functionalities of BRC20. For instance, ARC20~\cite{arc20tokens2024} removes the constraints of upper bound characters, while Rune~\cite{luminex2024rune} embeds more token data. (ii) Several studies focus on improving the performance and properties of specific components, such as off-chain indexers for data availability~\cite{wen2024modular}. (iii) Many UTXO-based blockchains draw inspiration from BRC20 on the Bitcoin network and begin to extend their inscription functions on-chain, such as DRC20~\cite{drc20standard2024} on Dogechain. (iv) Similarly, a lot of heterogeneous chains compared to Bitcoin also propose their inscriptions, such as Ethscriptions~\cite{ethscriptions2024introduction} on Ethereum, BSC20~\cite{bnbchaininscriptions2024} on BNB Smart Chain, ASC20~\cite{avaxmarket2024overview} on Avalanche, PRC20~\cite{prc20contract2024} on Polygon, SOLs~\cite{magicedenunlocking2024} on Solana, MRC20~\cite{coinlivemove2024} on Move, and APT20~\cite{apt20protocol2024} on Aptos. (v) Some studies focus on establishing connections between Bitcoin networks (inscribed tokens included) and wider blockchain ecosystems, such as relying on checkpoints~\cite{babylon} and bridges~\cite{wang2024bridging}.  (vi) Recent studies extended existing inscription mechanisms (which add plaintext scripts) by embedding more complex inscriptions, such as comments (e.g., NANDGate~\cite{bitvm}) and customized functions~\cite{lovejoy2023hamilton}. (vii) More broadly, many active teams invest efforts in implementing smart contracts on the Bitcoin network, also known as the Bitcoin layer two (L2)~\cite{qi2024sok} (more sharply focused than the previous L2~\cite{gudgeon2020sok}), such as BitVM~\cite{bitvm} and BitLayer~\cite{bitlayer}.

\smallskip
\noindent\textbf{BRC20 wallets are more than just wallets!}
Blockchain wallets are important in managing digital assets, performing tasks such as storing private keys, tracking balances, and facilitating transactions~\cite{houy2023security}\cite{mangipudi2023uncovering}\cite{eyal2022cryptocurrency}. However, unlike state-based wallets (e.g., MetaMask) that serve as an insulated interface~\cite{chatzigiannis2022sok}, BRC20 wallets offer expanded functionalities: (i) managing UTXOs that hold BRC20 tokens, which includes operations such as splitting and merging UTXOs based on transfer scripts; and (ii) interpreting and executing protocol-defined rules and functions encoded within transaction scripts (recall Sec.\ref{sec-brc20} for technical details), such as Ordinals, which govern transaction behavior. These wallets essentially function as off-chain executors (i.e., \textit{indexers}~\cite{binance3}), performing predefined functions to implement an off-chain state machine~\cite{wen2024modular}\cite{wang2024bridging}. 

\smallskip
\noindent\textbf{Attacking real blockchain projects.} Many studies present attacks against active blockchain projects with responsible disclosure, including  Ethereum~\cite{qin2023blockchain,zhang2024max,yaish2024speculative}, Binance~\cite{li2025binance}, Solana~\cite{smolka2023fuzz}, Monero~\cite{shi2025eclipse}, and PoA implementations~\cite{zhang2023time,ekparinya2020attack}. We commend their contributions to advancing blockchain security. 

\subsection{More Explanation on Ordinals and Inscriptions}

The Bitcoin Ordinals protocol introduces a mechanism for assigning unique serial numbers, referred to as \textit{ordinals}, to individual satoshis, the smallest units of Bitcoin. These ordinals can then be inscribed with arbitrary data such as text, images, or metadata, effectively transforming ordinary satoshis into immutable carriers of digital content.

The inscription process typically involves two separate Bitcoin transactions (referred to as \textit{\textbf{Tx1}}  and \textit{\textbf{Tx2}}  in this paper). The two transactions together fulfil what is informally known as the \textit{commit–reveal} pattern:

\begin{packeditemize}
\item \textbf{\textit{Tx1} (commitment transaction):} \textit{\textbf{Tx1}} creates a Taproot output that commits to the content of the inscription. It does so by including a specially constructed Taproot script leaf that embeds the hash of the full inscription data. The script uses a specific opcode pattern (e.g., $\mathsf{OP\_FALSE}$, $\mathsf{OP\_IF}$, $\ldots$, $\mathsf{OP\_ENDIF}$) to anchor the commitment. However, the actual content is not disclosed in \texttt{tx1}; only a cryptographic commitment is recorded on-chain.

\item \textbf{\textit{Tx2} (reveal transaction):} \textit{\textbf{Tx2}} spends the Taproot output created by \texttt{tx1}. In doing so, it provides the full inscription data inside the Taproot witness field. The data is wrapped in a standard envelope format with a MIME type identifier (e.g., $\mathsf{text/plain}$, $\mathsf{image/png}$), enabling interpretation by clients and inscription indexers. The Taproot spend path ensures that the content is fully verifiable and permanently recorded in Bitcoin’s UTXO set.

\end{packeditemize}

Once both \textbf{\textit{Tx1}} and \textbf{\textit{Tx2}} are confirmed, the inscription becomes an immutable on-chain object, anchored to a specific ordinal (i.e., a tracked satoshi). The protocol uses a deterministic first-in-first-out (FIFO) model to trace the movement of ordinals through UTXO creation and spending. As the inscribed satoshi is transferred, its content and identity persist, even across wallets and owners, without requiring changes to Bitcoin’s consensus rules.

Here, Ordinal tracking is purely client-side and does not affect consensus validation. Each inscription-aware wallet or indexer independently replays the blockchain from the genesis block to maintain the ordinal–satoshi mapping. This introduces soft guarantees about inscription integrity but avoids any protocol-level fork or rule change. 

\subsection{Deferred Attacking Details (cf. Sec.\ref{subsec-exeAttack})}
\label{app:attack-details}


We provide the deferred details of the 1st (fail), 3rd (\textcolor{purple}{success}), and 4th (\textcolor{purple}{success}) attack transactions, as well as the 1st recovery transaction associated with the 4th attack.

\begin{table}[h]
\centering

\renewcommand{\arraystretch}{0.85}
\resizebox{\linewidth}{!}{
    \begin{tabular}{|ll}
    \textbf{Attack} & \textcolor{black}{\textbf{\textit{1st Tx}}}\\
    Block ID & \#820,107 \\
    Inscription ID & \hlhref{https://ordiscan.com/inscription/46374444}{\textcolor{purple}{\#46374444}} \\
    Transaction ID & 8f609c036686f478b6123...291f27479fa9995 \\
    Source Address & bc1p7k2nuhqg5anj6ds5...e778x8swgk439 (Taproot address) \\
    Target Address & bc1qhuv3dhpnm0...sd0uhfdu7d (Binance's ORDI address) \\
    Amount & Input: 0.00028346 BTC, Output: 0.00000546 BTC \\
    Fee rate: & 200 sats/vB \\
    Fee used: & 27,800 sats (\textcolor{purple}{between} fees for \textit{\textbf{Tx1}} and \textit{\textbf{Tx2}}) \\
    Attack result & Fail \\

    &\\

    \textbf{Attack}  & \textcolor{black}{\textbf{\textit{3rd Tx}}} \\
    Block ID & \#820,123 \\
    Inscription ID & \hlhref{https://ordiscan.com/inscription/46385830}{\textcolor{purple}{\#46389818}} \\
    Transaction ID & b1de4af2a262bf2732f550...750c33e32a32c129e9486 \\
    Source Address & bc1ppre27yktdsz34...240c2g504qs403sks (Taproot address) \\
    Target Address & bc1qhuv3dhpnm0...sd0uhfdu7d (Binance's ORDI address)   \\
    Amount & Input: 0.00028346 BTC, Output: 0.00000546 BTC  \\
    Fee rate: & 201 sats/vB \\
    Fee used: & 27,800 sats \\
    Attack result & \textcolor{purple}{Success} \\
    
    & \\

    \textbf{Attack} & \textcolor{black}{\textbf{\textit{4th Tx}}} \\
    Block ID & \#820,125 \\
    Inscription ID & \hlhref{https://ordiscan.com/inscription/46385830}{\textcolor{purple}{\#46393039}} \\
    Transaction ID & fa5680d61e17dde63954...1c149164bbf2e35d0a4b3e1 \\
    Source Address & bc1pymxt2kuqrcm...l6uedrl28nsv6nu29 (Taproot address)  \\
    Target Address & bc1qhuv3dhpnm0...sd0uhfdu7d (Binance's ORDI address)  \\
    Amount & Input: 0.00014346 BTC, Output: 0.00000546 BTC  \\
    Fee rate: & 201 sats/vB \\
    Fee used: & 27,800 sats \\
    Attack result & \textcolor{purple}{Success} \\
    
    \end{tabular} 
}

\end{table}

We issued three recovery transactions to reactivate all dormant liquidity, and present a representative example.

\begin{table}[!h]
\centering

\renewcommand{\arraystretch}{0.8}
\resizebox{\linewidth}{!}{
    \begin{tabular}{|ll}
    \textbf{Recovery} & \textcolor{black}{\textbf{\textit{1st Tx}}} (corresponding to our \textcolor{purple}{\textit{4th attack}}) \\
    Block ID & \#820,133 \\
    Inscription ID & \hlhref{https://ordiscan.com/inscription/46393039}{\textcolor{teal}{\#46393039}} \\
    Transaction ID & fa5680d61e17dde63954c...c149164bbf2e35d0a4b3e1 \\
    Source Address & bc1qhuv3dhpnm0...sd0uhfdu7d (Binance's ORDI address) \\
    Target Address & bc1qhuv3dhpnm0...sd0uhfdu7d (same as source address)  \\
    Amount & Input 0.97875162 BTC, Output: 0.97736318 BTC \\
    Fee rate: & 404 sats/vB \\
    Fee used: & 138,844 sats (\textcolor{teal}{larger than} fees for \textit{\textbf{Tx2}})\\
    Recovery result & \textcolor{teal}{Success} \\
     
    \end{tabular}
}

\end{table}

\begin{table*}[!t]
  \centering
  \caption{\textbf{Applicability to mainstream BRC20 tokens and more similar tokens} (recall Sec.\ref{subsec-applicable}): We found that virtually all high-volume BRC20 assets on Bitcoin (e.g., \emph{Ordi}) inherit the same two-phase transfer logic and are thus within the attack surface. By contrast, inscription variants that fold commit + reveal into a single transaction or adopt a pure account model largely avoid the vulnerability.  }  \label{tab:brc-tokens}
   
\renewcommand{\arraystretch}{1} 

  \begin{threeparttable}
 \resizebox{\linewidth}{!}{
  \begin{tabular}{c|c c c c| ccc cc| c}
    \toprule
   \multirow{1}{*}{\textbf{Tokens}} & \multirow{1}{*}{\textbf{MarketCap}} & \multirow{1}{*}{\textbf{Price}} & \multirow{1}{*}{\textbf{T. Sup.}} & \multirow{1}{*}{\textbf{C. Sup. P.}} & \multirow{1}{*}{\textbf{Network}} & \multirow{1}{*}{\textbf{Protocol}} & \multirow{1}{*}{\textbf{Inscription}} & \multirow{1}{*}{\textbf{Transfer}}  & \multirow{1}{*}{\textbf{Required Txs}} & \multirow{1}{*}{\textbf{Applicable?}}  \\

   \midrule
   
    \hlhref{https://coinmarketcap.com/currencies/ordi/}{Ordi} & \$711M & \$33.52  & 21M & 100\%  & Bitcoin & BRC20 & Tx's $\mathsf{opcode}$ & UTXO/FIFO & Two ``bundled'' transactions &  \cmark  \\
   
   \hlhref{https://coinmarketcap.com/currencies/sats/}
   {Sats} & \$612M &  \$0.0002904 & 2.1Q  & 100\% & Bitcoin & BRC20 & Tx's $\mathsf{opcode}$ & UTXO/FIFO & Two ``bundled'' transactions  &  \cmark \\
   
   \hlhref{https://coinmarketcap.com/currencies/rats-ordinals/}{Rats} & \$115M & \$0.00011571  & 1T & 100\% & Bitcoin & BRC20 & Tx's $\mathsf{opcode}$ & UTXO/FIFO & Two ``bundled'' transactions  &  \cmark \\

    \hlhref{https://coinmarketcap.com/currencies/lever}
    {LeverFi} & \$74M &  \$0.002241 & 35B & 95.12\% & Bitcoin & BRC20 & Tx's $\mathsf{opcode}$ & UTXO/FIFO & Two ``bundled'' transactions &  \cmark \\
    
    \hlhref{https://coinmarketcap.com/currencies/pizza/}{PIZZA} & \$61M & \$2.9  & 21M & - & Bitcoin & BRC20  & Tx's $\mathsf{opcode}$ & UTXO/FIFO & Two ``bundled'' transactions  &  \cmark \\

    \hlhref{https://coinank.com/ordinals/brc20/WZRD}{WZRD} & \$57M & \$2.72  & 21M & 100\% & Bitcoin & BRC20 & Tx's $\mathsf{opcode}$ & UTXO/FIFO & Two ``bundled'' transactions   &  \cmark \\

    \hlhref{https://coinmarketcap.com/currencies/pups-ordinals/}{PUPS} & \$48M & \$6.19  & 7.77M & 100\% & Bitcoin & BRC20 & Tx's $\mathsf{opcode}$ & UTXO/FIFO & Two ``bundled'' transactions  &  \cmark \\
    
    \hlhref{https://coinank.com/ordinals/brc20/TEXO}{TEXO} & \$46M & \$0.09176 & 510M & 100\% & Bitcoin & BRC20 & Tx's $\mathsf{opcode}$ & UTXO/FIFO & Two ``bundled'' transactions  &  \cmark \\

    \hlhref{https://coinmarketcap.com/currencies/multibit/}{Multibit} & \$33M & \$0.03426 & 1B &  95\% & Bitcoin & BRC20 &  Tx's $\mathsf{opcode}$ & UTXO/FIFO & Two ``bundled'' transactions &  \cmark \\

    \hlhref{https://coinmarketcap.com/currencies/trac/}{TRAC} & \$28M & \$1.35  & 21M & 100\% & Bitcoin & BRC20 & Tx's $\mathsf{opcode}$ & UTXO/FIFO & Two ``bundled'' transactions   &  \cmark \\
    \hlhref{https://www.coingecko.com/en/coins/atomicals}{Atomicals} & \$27M & \$1.32 & 21M & - & Atomicals (BL2) & ARC20 & Tx's $\mathsf{metadata}$ & UTXO/FIFO & Two ``bundled'' transactions  & \xmark \\
    
    \hlhref{https://coinmarketcap.com/currencies/piin/}{Piin} & \$11.27M & \$0.0001129 & 100B & 100\% & Bitcoin & BRC20 & Tx's $\mathsf{opcode}$ & UTXO/FIFO & Two ``bundled'' transactions  &  \cmark \\
    \hlhref{https://coinmarketcap.com/currencies/orange-crypto/}{Orange} & \$8M & \$0.1002 & 100M & 25.66\% & Bitcoin & BRC20 & Tx's $\mathsf{opcode}$ & UTXO/FIFO & Two ``bundled'' transactions  &  \cmark \\
    
    \hlhref{https://coinmarketcap.com/currencies/bnsx-ordinals/}{BNSx} & \$1.22M & \$0.0565  & 21M & 38.29\% & Bitcoin & BRC20 & Tx's $\mathsf{opcode}$ & UTXO/FIFO & Two ``bundled'' transactions   &  \cmark \\

    \hlhref{https://coinmarketcap.com/currencies/cats-ordinals/}{cats} & \$1.19M & \$0.0005659
    & 2.1B & 77.79\% & Bitcoin & BRC20 & Tx's $\mathsf{opcode}$ & UTXO/FIFO & Two ``bundled'' transactions   &  \cmark \\
    
   \midrule

    \hlhref{https://coinmarketcap.com/currencies/ethscriptions/}{ETHS} & \$19M & \$0.9251 & 21M & 100\% & Ethereum & ETHS20 & Tx's $\mathsf{calldata}$  & Account model & Single transaction &\xmark \\


   \hlhref{https://www.coingecko.com/en/coins/dogi}{DOGI} & \$17M & \$0.8198 & 21M & 100\% & Dogecoin & DRC20 & Tx’s $\mathsf{opcode}$ & UTXO/FIFO & Two “bundled” transactions  & \cmark \\


    \hlhref{https://coinmarketcap.com/currencies/gram/}{Gram} & \$9.89M & \$0.004007 & 5B & 100\% & Arbitrum & TON20 & Tx’s $\mathsf{opcode}$ & Account model & Single transaction & \xmark \\

    \hlhref{https://www.coingecko.com/en/coins/avav-asc-20}{BSCS} & \$1.15M & \$0.004677 & 500M & 79.8\% & BNB Chain & BSC20 & Tx's $\mathsf{calldata}$ & Account model & Single transaction & \xmark \\

    \hlhref{https://coinmarketcap.com/currencies/sols/}{SOLS} & \$90,278 & \$0.004299 & 21M & 100\% & Solana & SPL20 & Tx's $\mathsf{calldata}$ & Account model & Single transaction & \xmark \\


    \bottomrule
  \end{tabular}
  }
  \smallskip
  \begin{tablenotes}
       \footnotesize
       \begin{minipage}{\textwidth}
       \item[] \textbf{Note}: The table includes active inscription-based tokens (with liquidity and market value). Tokens like \hlhref{https://www.coingecko.com/en/coins/quantum-pipeline}{PIPE}, hot but now without values, are excluded.
       \item[]  \textbf{T. Sup.} for total supply; \textbf{C. Sup. P.} for circulating supply percentage; \textbf{FIFO} for First-in-first-out; \textbf{BL2} for Bitcoin Layer two; \textbf{-} for data unavailable.
      \end{minipage}
     \end{tablenotes}
  \end{threeparttable}
\end{table*}


\subsection{Operation Tolerance Calculation} 
\label{appendix:tolerance}

\noindent\textbf{Operational tolerance of Binance}
By applying typical values to our formula, we demonstrate the calculation of the operational tolerance range \(T_{\text{bar}}\). For this analysis, we consider Binance as our target CEX.

Due to the lack of direct data from Binance, we approximate its values by analyzing ORDI transaction data from 2024 on OKX~\cite{okx24}, a comparable CEX with a slightly smaller user base than Binance. We derive the following parameters\footnote{The specific values are illustrative estimates based on typical CEXs and publicly available information up to Sep. 2024. Actual values may vary depending on the exchange’s size, involved tokens, and market conditions.}: \( L_{\text{avail}}^\text{high} \), \( L_{\text{avail}}^\text{low} \) is 5,000,000 and 2,500,000 tokens, respectively; \(L_{\text{req}}\) is estimated at 2,000,000 tokens; and \(V\) is around 1,000,000 tokens per hour. Based on these values, we calculate \(T_{\text{bar}}^h\) to be \textit{3 hours} as the upper bound. Conversely, the lower bound $T_{\text{bar}}^l$ is calculated to be \textit{0.5 hours}.

\smallskip
\noindent\textbf{Operational tolerance range.} By calculation, our attack targets CEXs, where a reasonable operational tolerance range is estimated to be \textbf{between 0.5 and 3 hours}, depending on their available liquidity reserves and transaction volumes. For DeFi protocols that provide liquidity pools or offer staking services, \( T_{\text{bar}} \) tends to be slightly higher, usually spanning from a few hours to up to 24 hours, due to their decentralized nature and generally lower immediate liquidity demands compared to centralized exchanges. On the other hand, wallet services or newly launched DeFi projects may exhibit a shorter \( T_{\text{bar}} \), often less than 30 minutes, as they have lower liquidity reserves and handle fewer transactions.

\subsection{Ethics Considerations}
\label{sec-ethics}


\noindent\textbf{Coordinated collaboration.}  
The mempool-level pinning vulnerability was found and analysed jointly with Binance’s researchers.  Every in-production test was executed only after written approval from Binance infrastructure leads, on a single ORDI hot wallet that both parties preselected for its high activity and routine liquidity buffers.

\smallskip
\noindent\textbf{Zero loss on user funds.}  
All tests were rate-limited and continuously monitored; no user balances were impacted (impact implies loss, NOT delay) during experiments (we do NOT count unrealized opportunity costs here) Following each controlled run, transactions were reversed as appropriate, and mempool states were promptly restored to baseline. 

It is worth noting (and confirmed through our coordination) that short-term suspension of specific wallet functions is a common operational pattern among infrastructure providers. For major system upgrades, services are typically labeled as “under maintenance” or “on hold.” However, for brief or backend adjustments, functional pauses are often handled silently, without public announcement. This silent buffering allows providers to make low-risk changes while maintaining service continuity—consistent with standard DevOps practices across large-scale platforms.

As a relatable example, when a user initiates a token withdrawal from a centralized exchange, the request typically enters an approval queue. The duration of this approval process varies across exchanges and time windows, effectively introducing a built-in buffer for backend adjustments or temporary functional halts. Our controlled tests were scoped within this operational envelope.

\smallskip
\noindent\textbf{Non-exploitation pledge.}  
  The authors derived no financial benefit, placed no trades, and introduced no artificial market pressure.  Results were withheld from third parties until Binance had confirmed mitigation readiness.

\smallskip
\noindent\textbf{Selective disclosure.}  
  Implementation-level scripts and automation parameters that could facilitate copycat attacks are deliberately omitted.  A complete technical report has been provided privately to Binance.

\smallskip
\noindent\textbf{Legal and compliance check.}  
  The study involved no personal data, no deanonymisation of users, and no actions contravening U.S./EU computer-misuse statutes.  Institutional review confirmed that formal IRB oversight was not required because no human subjects were involved.

\end{document}